\documentclass[aps,pra,twocolumn,letterpaper,showpacs,superscriptaddress,floatfix,10pt]{revtex4-1}
\usepackage[pdftex]{graphicx}% Include figure files
\graphicspath{ {./images/} }
\usepackage{dcolumn}% Align table columns on decimal point
\usepackage{bm}% bold math
\usepackage{bbm}% bold math
\usepackage[usenames, dvipsnames]{color}
\usepackage{comment}
\usepackage[colorlinks, linkcolor=blue]{hyperref}

\usepackage{layouts}

\usepackage[T1]{fontenc}

\usepackage{amsfonts}
\usepackage{amssymb}
\usepackage{amsmath}

\usepackage{amsthm}
\theoremstyle{remark}

\usepackage{multirow}
\usepackage[scr=boondoxo, scrscaled=1.05]{mathalfa}

\usepackage{subfigure}
\usepackage{overpic}

\usepackage{array}
\newcolumntype{L}[1]{>{\raggedright\let\newline\\\arraybackslash\hspace{0pt}}m{#1}}
\newcolumntype{C}[1]{>{\centering\let\newline\\\arraybackslash\hspace{0pt}}m{#1}}
\newcolumntype{R}[1]{>{\raggedleft\let\newline\\\arraybackslash\hspace{0pt}}m{#1}}
\usepackage{tabularx}

\def\KeyWord#1{$\backslash$\IfColor{$\!\!$\textRed{#1}\textBlack}{#1}$\!\!$}

\usepackage{wasysym}

\newcommand{\integers}{\mathbb{Z}}

\renewcommand{\d}{\mathrm{d}}

\newcommand{\Bn}{{\vec{n}}}
\newcommand{\Bm}{{\vec{m}}}

\newcommand{\BO}{{\vec{\Omega}}}
\newcommand{\Bt}{{\vec{\theta}}}

\newcommand{\Ptop}{{P_{\mathrm{top}}}}

\def\bra#1{\langle#1|}
\def\ket#1{|#1\rangle}

\def\qexp#1#2{\bra{#2}#1\ket{#2}}

\def\tr#1{\mathrm{Tr}\left[#1\right]}

\begin{document}

\title{Coupled Layer Construction for Synthetic Hall Effects in Driven Systems}

\author{David M. Long}
\email{dmlong@bu.edu}
\affiliation{Department of Physics, Boston University, Boston, Massachusetts 02215, USA}

\author{Philip J. D. Crowley}
\affiliation{Department of Physics, Massachusetts Institute of Technology, Cambridge, Massachusetts 02139, USA}

\author{Anushya Chandran}
\affiliation{Department of Physics, Boston University, Boston, Massachusetts 02215, USA}

\date{\today}

\begin{abstract}
Quasiperiodically driven fermionic systems can support topological phases not realized in equilibrium.
The fermions are localized in the bulk, but support quantized energy currents at the edge.
These phases were discovered through an abstract classification, and few microscopic models exist.
We develop a coupled layer construction for tight-binding models of these phases in \(d\in\{1,2\}\) spatial dimensions, with any number of incommensurate drive frequencies \(D\).
The models exhibit quantized responses associated with synthetic two- and four-dimensional quantum Hall effects in the steady state.
A numerical study of the phase diagram for \((d+D) = (1+2)\) shows: (i) robust topological and trivial phases separated by a sharp phase transition; (ii) charge diffusion and a half-integer energy current between the drives at the transition; and (iii) a long-lived topological energy current which remains present when weak interactions are introduced.
\end{abstract}

\maketitle

\section{Introduction}
    \label{sec:intro}

    Time dependent driving is a useful tool in quantum simulation and computation, and can create new nonequilibrium phases of matter~\cite{Bukov2015,Eckardt2017,Gross2017,Oka2019,Rudner2020}. The most studied example is periodic (Floquet) driving. Beyond the ubiquitous Rabi effect, Floquet driving is an emerging tool in band~\cite{Oka2009,Bukov2015,Eckardt2017,Gross2017,Oka2019} and crystal lattice engineering~\cite{Disa2021}. In the steady state, Floquet phases can also exhibit period doubling~\cite{Khemani2016,Else2016,Khemani2019,Else2020} and topological phases~\cite{Blackburn2005,Kitagawa2010,Lindner2011,Rudner2013,Rechtsman2013,Nathan2015,Titum2016,Nathan2017,Roy2017a,Roy2017,Kolodrubetz2018,Maczewsky2017,Li2018,Schuster2019,Wintersperger2020}, some of which are impossible in equilibrium.

    \begin{figure}[h!]
        \centering
        \includegraphics[width=\linewidth]{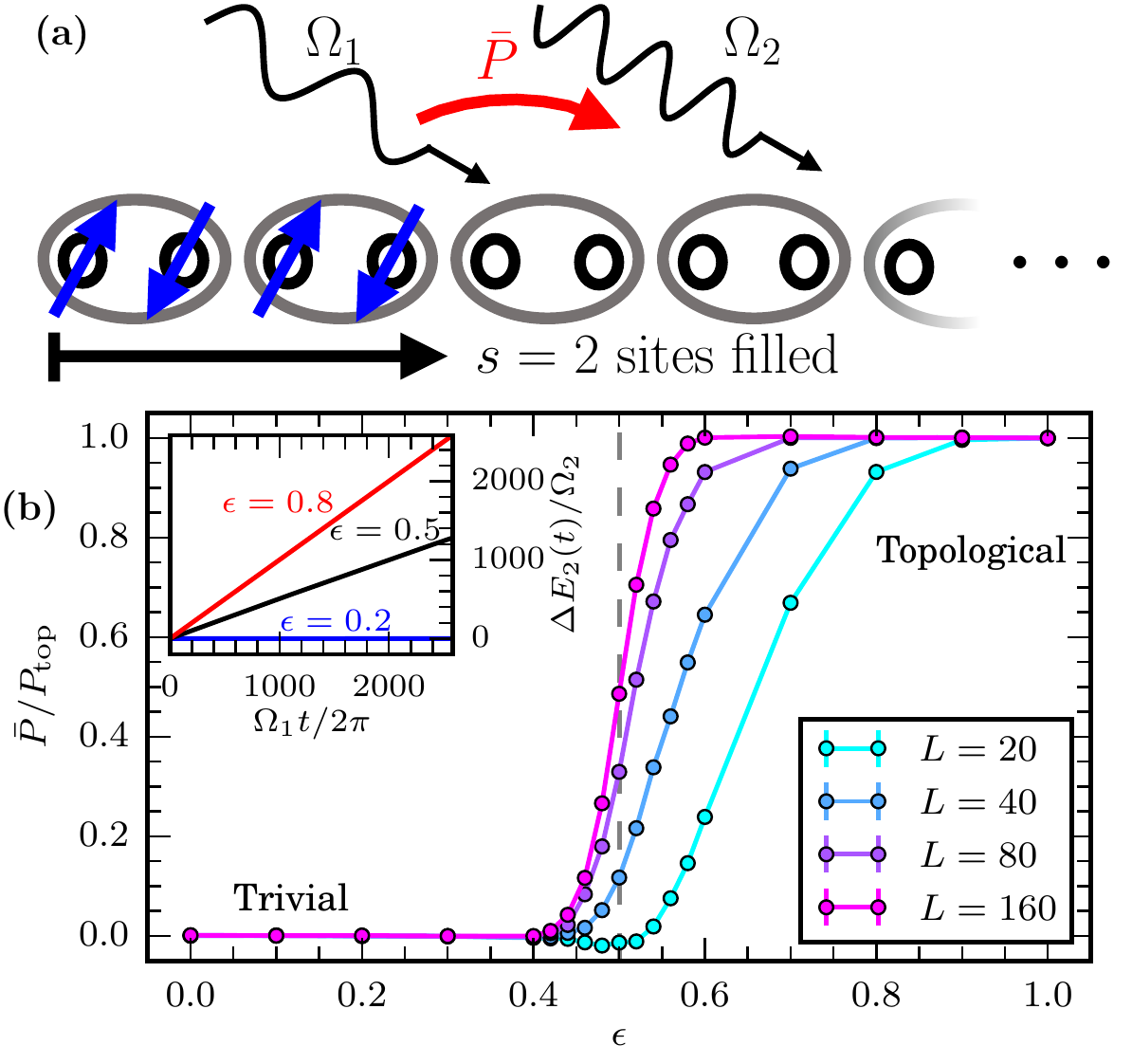}
        \caption{(\textbf{a}) The quasiperiodic Floquet-Thouless energy pump (QP pump) is a nonequilibrium phase of disordered fermionic chains driven by incommensurate frequencies \(\Omega_1\) and \(\Omega_2\). When sufficiently many sites (\(s\)) from one edge are filled, the chain mediates a topologically quantized average energy current between the drives, \(\bar{P} = P_{\mathrm{top}} W\) (\(W \in \integers\)). (\textbf{b}) The parameter \(\epsilon\) deforms a coupled layer model of (\textbf{a}) from the trivial phase (average energy current \(\bar{P} = 0\)) to the topological phase (\(\bar{P} = \Ptop\)). The transition between the phases sharpens with increasing chain length \(L\). At the critical point (\(\epsilon=1/2\)), the pumping rate is half the topological value. (Inset) The total energy pumped into drive 2, \(\Delta E_2\), is linear in time in both phases and at the transition. \emph{Parameters in model~\eqref{eqn:H_QP pump}}: \(s = L/4\), \(\bar{P}\) is averaged over 200 disorder configurations and initial phases, and as in \autoref{fig:corr_len}.}
        \label{fig:pump_PD}
    \end{figure}

    For instance, two-dimensional fermionic systems support a topological phase with chiral current-carrying edge modes~\cite{Kitagawa2010,Rudner2013,Nathan2015,Titum2016,Nathan2017,Maczewsky2017,Wintersperger2020}. Such behavior is only possible in a two-dimensional static insulator when populated bands have a nontrivial Chern number, but this requirement is evaded in the driven setting. The drive is engineered to move fermions in short loops---similar to cyclotron orbits---which results in skipping orbits at the edge of the system, while the bulk may be fully localized. These \emph{anomalous Floquet-Anderson insulators} (AFAIs) have prompted keen interest in the nonadiabatic properties of driven phases of matter~\cite{Roy2017a,Roy2017,Maczewsky2017,Peng2018a,Rudner2020}.

    The study of quasiperiodically driven systems---those driven by multiple drives of incommensurate frequencies---goes back several decades~\cite{Ho1983,Luck1988,Casati1989,Jauslin1991,Blekher1992,Jorba1992,Feudel1995,Bambusi2001,Gentile2003,Chu2004,Gommers2006,Chabe2008,Cubero2018,Nandy2018,Ray2019}. The characterization of novel effects in these systems as being properties of new nonequilibrium phases of matter is more recent~\cite{Nandy2017,Dumitrescu2018,Kolodrubetz2018,Peng2018,Peng2018b,Zhao2019,Else2019,Friedman2020,Crowley2020,Boyers2020,Long2021,Nathan2020b,Long2022b,Dumitrescu2022}. Quasiperiodic driving can stabilize topological edge modes in interacting chains without any symmetry~\cite{Friedman2020,Dumitrescu2022}, and even driven qubits can exhibit quantized responses~\cite{Martin2017,Crowley2019,Crowley2020,Boyers2020,Long2021,Nathan2020b}.

    There is an intimate connection between quasiperiodically driven topological phases and Floquet phases in higher dimensions~\cite{Long2021,Nathan2020b}. Specifically, the steady states of a \(d\)-dimensional tight-binding model driven by \(D\) incommensurate tones follow from the eigenstates of a \((d+D)\)-dimensional static model on a \emph{frequency lattice}~\cite{Sambe1973,Ho1983,Verdeny2016,Martin2017}\footnote{The frequency lattice models do not represent physical lattice models, as their spectrum is unbounded from above and below, but see Ref.~\cite{Lapierre2021}.}. This construction relates the AFAI with \((d+D) = (2+1)\) to a \((1+2)\) \emph{quasiperiodic Floquet-Thouless energy pump} (QP pump)---a one-dimensional phase of two-tone-driven fermions. This connection reveals that the QP pump supports localized edge modes which mediate an energy current between the drives (\autoref{fig:pump_PD}). This energy current has a quantized average value,
    \begin{equation}
        \bar{P} = \Ptop W,
        \quad\text{where}\quad
        \Ptop = \frac{\Omega_1 \Omega_2}{2\pi},
        \label{eqn:Ptop}
    \end{equation}
    and \(W\in\integers\) is a winding number invariant. Further, the QP pump has remarkable coherence properties which allow for the preparation of highly excited nonclassical states in quantum cavities~\cite{Long2022b}.

    Reference~\cite{Long2021} classified localized phases with any \((d+D)\). When \((d+D)\) is odd there is an integer classification of \emph{anomalous localized topological phases} (ALTPs). However, the abstract classification does not reveal observable properties of these phases. It is thus useful to have simple models for each ALTP. Such models would also guide experimental realizations of these phases.

    In this paper, we devise a coupled layer construction for any \((1+D)\)-dimensional ALTP. We demonstrate the construction in detail for the simplest example of the QP pump (\autoref{sec:model}). Exploiting the mapping to the frequency lattice, we show the QP pump can be constructed from layers supporting delocalized chiral modes, just as in familiar integer quantum Hall phases~\cite{Klitzing1980,Laughlin1981,Thouless1982,Sondhi2001,Kane2002}. The layers for the QP pump are fermionic sites, finely tuned to support pumping modes with equal and opposite average energy currents between the drives. These pumping modes can be coupled in one of two ways: within a site, resulting in a trivial phase; or between sites, resulting in a topological phase with dangling edge modes (\autoref{fig:wires}).

    The coupled layer construction can also be adapted to produce a \((2+3)\)-dimensional ALTP with edge states exhibiting a synthetic four-dimensional quantum Hall effect (\autoref{sec:general_model})~\cite{Lohse2018}. The physical response is an energy current between two of the drives supported at one of the (one-dimensional) edges,
    \begin{equation}
        \bar{P} = \frac{\Omega_3 \Omega_4}{(2\pi)^2} B L_y W + O(B^2).
        \label{eqn:P4DQHE_intro}
    \end{equation}
    Here, \(B\) is a synthetic magnetic field, \(L_y\) is the linear dimension of the pumping edge, and \(W \in \integers\) is a winding number.

    We numerically investigate the QP pump coupled layer model, and obtain the phase diagram shown in \autoref{fig:pump_PD}(b) as a function of the interlayer coupling strength \(\epsilon\). The model has two localized phases---one topological and one trivial (\autoref{subsec:loc_phase})---separated by an isolated critical point (\autoref{subsec:crit_point}). The critical point exhibits a half-integer energy current, \(\bar{P} = \Ptop/2\), with critical exponents suggestive of the two-dimensional integer quantum Hall universality class. In the topological phase, the energy current is very robust to weak interactions (\autoref{sec:interactions}). It persists for an extremely long time, even when interactions cause the system to ultimately thermalize.

\section{Background}
    \label{sec:background}

    ALTPs are characterized through their back action on the drives~\cite{Martin2017,Kolodrubetz2018,Crowley2019,Nathan2019c,Long2021,Nathan2020b}. The \emph{frequency lattice} formalism (\autoref{subsec:frq_lat}) facilitates a description of the steady states of the system, together with the drive. This formalism reveals the properties of the QP pump straightforwardly (\autoref{subsec:QP-pump}).

    \subsection{The frequency lattice}
        \label{subsec:frq_lat}

        Quasiperiodically driven systems may be mapped to lattice problems with additional synthetic dimensions---one for each incommensurate tone~\cite{Sambe1973,Ho1983,Blekher1992,Verdeny2016,Martin2017}. We review this construction below.

        The models we consider are fermionic tight-binding models driven by incommensurate periodic tones with frequencies \(\Omega_j\) (\(j \in \{1,\ldots, D\}\)). It is useful to write the time-dependent Hamiltonian as being a function of the drive phases \(\theta_j(t) = \Omega_j t + \theta_{0j}\), which for brevity we assemble into a vector,
        \begin{equation}
            H(t) = H(\Bt_t), \quad \Bt_t = \sum_j \theta_j(t) \hat{e}_j,
            \label{eqn:H_theta}
        \end{equation}
        where \(\hat{e}_j\) form a basis of unit vectors. The Hamiltonian may be either single particle or many body. In the coupled layer models we will use a many-body notation for \(H\) (in terms of creation and annihilation operators), though the model is quadratic.

        To discuss topology in the context of quasiperiodically driven systems, we must have an appropriate notion of steady states. These are the \emph{quasienergy states}, \(\ket{\psi_\alpha(t)}\). When they exist, they form a complete basis of solutions to the Schr\"odinger equation, and have the special form (\(\hbar = 1\))
        \begin{equation}
            \ket{\psi_\alpha(t)} = e^{-i \epsilon_\alpha t} \ket{\phi_\alpha(\Bt_t)},
            \label{eqn:quasi_states}
        \end{equation}
        where \(\epsilon_\alpha\) is called the \emph{quasienergy} and \(\ket{\phi_\alpha(\Bt_t)}\) is a smooth, periodic function of each \(\theta_j\), and thus is naturally defined on a torus of drive phases. We will also refer to \(\ket{\phi_\alpha(\Bt_t)}\) as a quasienergy state. Unlike in periodically driven systems (\(D=1\)), where Floquet's theorem~\cite{Floquet1883} guarantees the existence of a complete set of quasienergy states, the quasienergy states~\eqref{eqn:quasi_states} need not always exist~\cite{Jauslin1991,Blekher1992,Crowley2019}.

        A formal rewriting of the eigenstate equation for the quasienergy states in terms of a \emph{frequency lattice} model assists in understanding the states~\cite{Sambe1973,Ho1983,Verdeny2016,Martin2017,Crowley2019}. The quasienergy states solve the Schr\"odinger equation,
        \begin{equation}
            (H(\Bt_t) - i \partial_t)\ket{\phi_\alpha(\Bt_t)} = \epsilon_\alpha \ket{\phi_\alpha(\Bt_t)}.
            \label{eqn:t_quasieqn}
        \end{equation}
        By Fourier transforming this equation with respect to each drive phase, and introducing auxiliary states \(\ket{\Bn}\) associated to the Fourier indices, the problem of finding \(\ket{\phi_\alpha(\Bt_t)}\) is reformulated as a lattice problem. Equation \eqref{eqn:t_quasieqn} is then recast as the eigenvalue equation
        \begin{equation}
            \tilde{K} \ket{\tilde{\phi}_\alpha} = \epsilon_\alpha \ket{\tilde{\phi}_\alpha}.
        \end{equation}
        Here, we have adopted a notation in which the lattice quasienergy states are given by
        \begin{equation}
             \ket{\tilde{\phi}_\alpha} = \sum_{\Bn \in \integers^D} \ket{\phi_{\alpha,\Bn}} \ket{\Bn}
        \end{equation} 
        (where \(\ket{\phi_{\alpha,\Bn}}\) are the Fourier components of \(\ket{\phi_\alpha(\Bt)}\)), and the frequency lattice Hamiltonian, called the quasienergy operator, is given by
        \begin{equation}
            \tilde{K} = \sum_{\Bn,\Bm \in \integers^D} (H_{\Bn-\Bm} e^{i(\Bn-\Bm)\cdot \Bt_0} - \Bn\cdot\BO \delta_{\Bn \Bm}) \ket{\Bn}\bra{\Bm}
            \label{eqn:frq_lat}
        \end{equation}
        (where \(H_\Bn\) are the Fourier components of \(H(\Bt)\)). Smooth quasiperiodic solutions to Eq.~\eqref{eqn:t_quasieqn} exist when \(\tilde{K}\) has localized eigenstates. The quasienergy states are then obtained as
        \begin{equation}
            \ket{\phi_\alpha(\Bt_t)} = \sum_{\Bn \in \integers^D} e^{-i \Bn \cdot \Bt_t} \ket{\phi_{\alpha,\Bn}}.
            \label{eqn:quasi_st_sum}
        \end{equation}

        When \(H\) is a single-particle Hamiltonian, the quasienergy operator \(\tilde{K}\) describes a \((d+D)\)-dimensional tight-binding model with a linear potential in the \(\BO = \sum_j \Omega_j \hat{e}_j\) direction. We refer to \(\BO\) as an electric field in the frequency lattice. When \(\Bn \cdot \BO  \neq 0\) for all \(\Bn \neq 0\), this electric field is incommensurate to the lattice. Further, the initial phase of the drive \(\Bt_0\) acts as a vector potential in the frequency lattice. Said vector potential is uniform in all the synthetic dimensions---modulation of \(\Bt_0\) in real space is necessary to produce a nontrivial flux through finite loops in frequency space~\cite{Peng2018b}.

        The frequency lattice model~\eqref{eqn:frq_lat} can be understood as the semiclassical limit of a cavity system. Replacing the classical drives in Eq.~\eqref{eqn:H_theta} with quantum cavities, the frequency lattice model~\eqref{eqn:frq_lat} is recovered when writing the model in the Fock state basis. The auxiliary frequency lattice states \(\ket{\Bn}\) can thus be interpreted as photon occupations for the drives. The potential \(\Bn\cdot\BO = \sum_j n_j \Omega_j\) accounts for the drive energies.

    \subsection{Quasiperiodic Floquet-Thouless Energy Pump}
        \label{subsec:QP-pump}

        While ALTPs have been classified in all dimensions, the best understood example with \(D>1\) is the QP pump~\cite{Kolodrubetz2018,Long2021,Nathan2020b}. The QP pump is the ALTP with spatial dimension \(d=1\) and two incommensurate drives, \(D = 2\) (the \((1+2)\)-dimensional ALTP). This section summarizes some known facts regarding this phase. (See Ref.~\cite{Lapierre2021} for the related \((3+0)\)-dimensional phase.)

        The bulk topological invariant associated to ALTPs is a winding number, \(W\). In the QP pump, the corresponding signature at the edge of the system is a quantized average pumping of energy between the drives. The direction in which this pumping proceeds is fixed by the sign of \(W\) and which edge is being considered.

        Localization of the quasienergy states in the synthetic dimensions is crucial here. Without this, one cannot define steady states as in Eq.~\eqref{eqn:quasi_st_sum}. In the frequency lattice tight-binding model~\eqref{eqn:frq_lat}, localization may occur due to the inhomogeneous potential \(\Bn\cdot\BO\). The topological classification of ALTPs further assumes localization in the spatial dimensions~\cite{Long2021,Nathan2020b,Lapierre2021}. The specific mechanism of localization---random or correlated spatial disorder, Stark localization through a linear potential, or otherwise---is unimportant.

        The observable which measures the rate of energy transfer into the second drive is
        \begin{equation}
            P(t) = -\Omega_2 \partial_{\theta_2} H(\Bt_t).
            \label{eqn:pow}
        \end{equation}
        Writing \(\rho_s\) for the Slater determinant state with the first \(s\) states from the edge filled (and potentially other states in the bulk), we have
        \begin{equation}
            \bar{P} := \lim_{T\to \infty} \frac{1}{T}\int_0^T \d t\, \tr{P(t) \rho_s(t)} \\
            = \frac{\Omega_1 \Omega_2}{2\pi} W ,
            \label{eqn:Pbar}
        \end{equation}
        where \(\rho_s(t)\) is the time evolved state from the initial state \(\rho_s\), and \(W \in \integers\) is the winding number~\cite{Kolodrubetz2018,Long2021,Nathan2020b}. We denote the topological pumping rate as
        \begin{equation}
            \bar{P} = \Ptop W + O(e^{-s/\zeta})
            \label{eqn:Pbar2}
        \end{equation}
        with \(\Ptop = \tfrac{\Omega_1 \Omega_2}{2\pi}\) as in Eq.~\eqref{eqn:Ptop}, and where \(\zeta\) is the single-particle localization length. (Equation~\eqref{eqn:Pbar2} holds for any initial phase and disorder realization which results in localization. However, the data we plot in \autoref{fig:pump_PD} and later figures include an average over initial phases \(\Bt_0\) and disorder. This reduces the \(O(T^{-1})\) noise due to calculating the average \(\bar{P}\) over a finite time \(T\).)

\begin{figure}
        \centering
        \includegraphics[width=\linewidth]{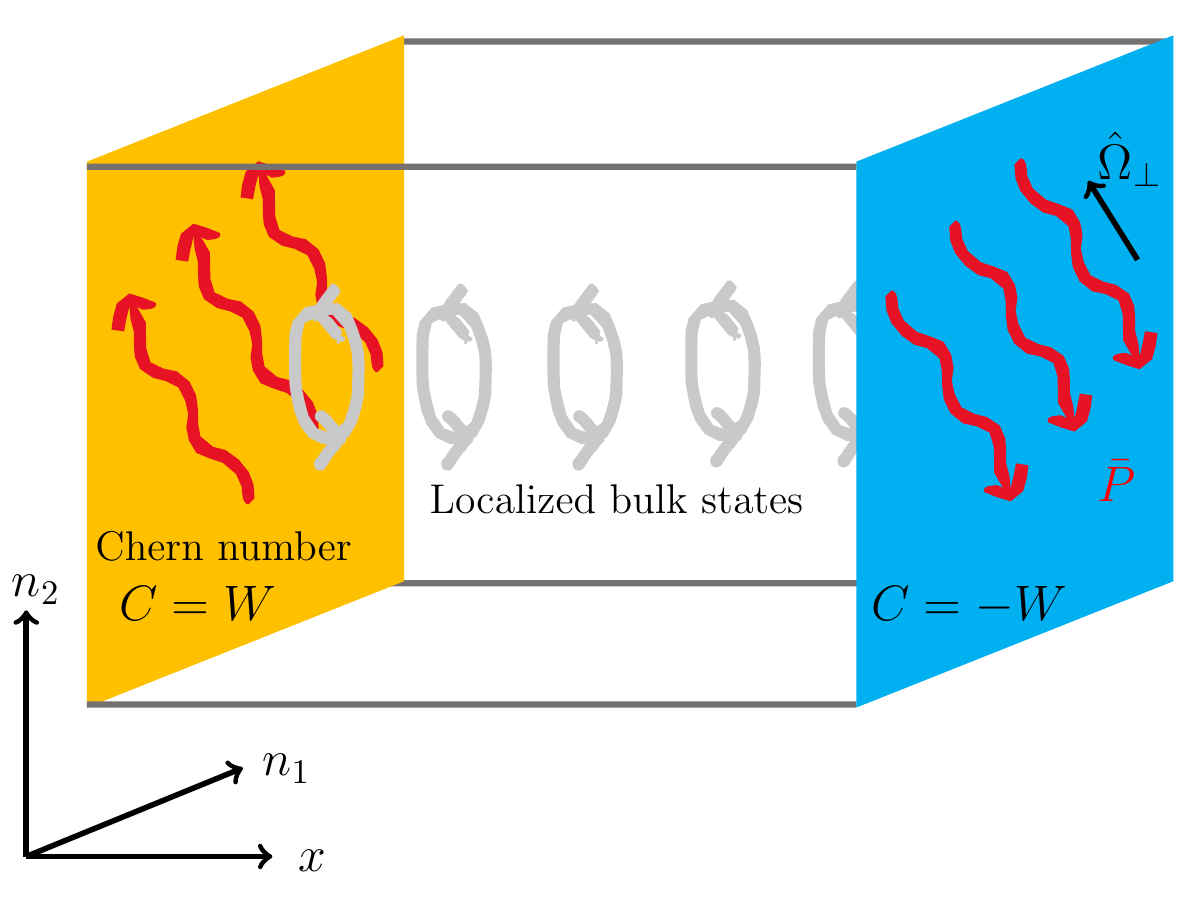}
        \caption{In the topological phase of the QP pump, the (single-particle) edge states have a Chern number in the frequency lattice. The Chern number \(C\) is given by the bulk winding number \(W\), up to a sign depending on which edge is considered. The frequency lattice electric field \(\BO\) induces a transverse current (along \(\hat{\Omega}_\perp\)) through the quantum Hall effect when one Chern state is completely filled. This is the energy pumping response of the edge, \(\bar{P}\). (See also Ref.~\cite{Lapierre2021}.)}
        \label{fig:Chernlayers}
    \end{figure}

        In terms of the frequency lattice, the pumping states correspond to delocalized edge states. A state initialized with photon numbers \(n_1\) and \(n_2\) can evolve into another with \(n_1 + \bar{P} t/\Omega_1\) and \(n_2 - \bar{P} t/\Omega_2\). As such, if \(\bar{P} \neq 0\), the eigenstates at the edge must be delocalized in the direction
        \begin{equation}
            \hat{\Omega}_\perp \propto \Omega_2 \hat{e}_1 - \Omega_1 \hat{e}_2.
        \end{equation}
        That is, perpendicular to \(\BO\). We will write \(n_\perp = \Bn \cdot \hat{\Omega}_\perp\) for the corresponding frequency lattice coordinate (Figs.~\ref{fig:wires} and \ref{fig:Chernlayers}).

        There is another observable which reveals the topology of the QP pump, in addition to the edge modes. This is a circulation of energy between the drives in the bulk~\cite{Long2021,Nathan2020b}. As we will not focus on this observable, a cartoon picture for it suffices. Fermions move right (say) then absorb a photon, move left, and emit a photon. This results in a small loop in the frequency lattice (\autoref{fig:wires}), and an observable associated to this motion turns out to have a quantized averaged expectation value proportional to the winding number \(W\).

\section{Coupled layer model for the QP pump}
    \label{sec:model}

\begin{figure}
        \centering
        \includegraphics[width=\linewidth]{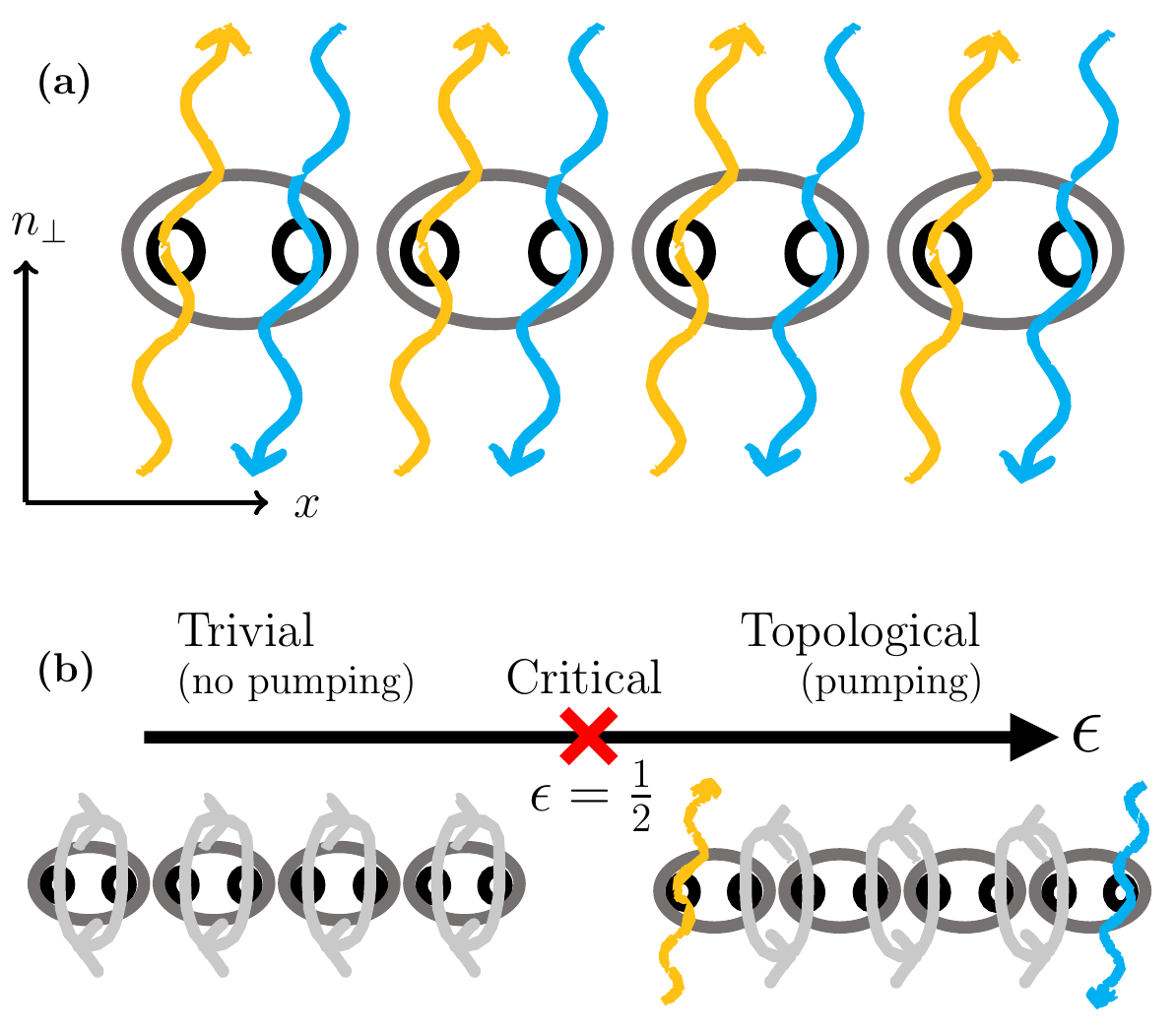}
        \caption{The QP pump may be constructed from a chain of sites in a coupled layer model. (\textbf{a}) The building blocks of the model are counter-diabatically driven spinful fermionic sites. The quasienergy states of a fermion on a site pump energy between the two drives at a quantized rate. In the frequency lattice, this is a current along the \(\hat{\Omega}_\perp\) direction, with coordinate \(n_\perp\) (\autoref{fig:Chernlayers}). (\textbf{b}) Coupling the sites causes the pumping modes to hybridize and localize. The tuning parameter \(\epsilon\) interpolates between a trivial pattern of hybridization, where all states are localized (\(\epsilon=0\)), and a topological one, where pumping modes remain at the edge (\(\epsilon=1\)). Finite-size scaling (\autoref{subsec:crit_point}) suggests the model has a single critical point at \(\epsilon=1/2\).}
        \label{fig:wires}
    \end{figure}

    The existence of delocalized edge states in the QP pump (\autoref{subsec:QP-pump}) suggests it may be possible to create a kind of coupled layer construction for this phase (\autoref{fig:wires}). By taking sites tuned to criticality---in the sense of having delocalized energy pumping modes (\autoref{subsec:sites})---and coupling them so as to either cancel all pumping or leave dangling edge modes (\autoref{subsec:coupling}), we can construct models of the trivial phase and of the QP pump, respectively.

    \subsection{Sites}
        \label{subsec:sites}

        Generic two-tone driven few-level systems are localized in the frequency lattice when \(|\BO| > 0\)~\cite{Crowley2019,Long2022a}. As such, they do not pump energy in the steady state. However, fine tuning in the form of an additional counterdiabatic drive can produce delocalized modes which support quantized energy pumping~\cite{Crowley2019}. Such finely tuned two-level systems will form the sites of the coupled layer construction.

        The model is defined in terms of a one-dimensional chain of spinful fermionic sites, with corresponding annihilation operators \(c_{x \mu}\), where \(x\) labels position and \(\mu\) is a spin index.

        Explicitly, the Hamiltonian with \(L\) sites is
        \begin{equation}
            H_0(\Bt) = \sum_{x=0}^{L-1} c^\dagger_{x \mu}\left[-(\vec{B}+\vec{B}_{\mathrm{CD}})\cdot\vec{\sigma}_{\mu \nu}/2\right] c_{x \nu},
            \label{eqn:H0}
        \end{equation}
        where \(\vec{\sigma}\) is a vector of Pauli matrices, summation over the spin indices is implied, and
        \begin{equation}
            \vec{B}(\Bt) = B_0 \left[\sin\theta_1 \mathbf{\hat{x}} + \sin\theta_2 \mathbf{\hat{y}}+ (1-\cos\theta_1-\cos\theta_2)\mathbf{\hat{z}}\right],
            \label{eqn:B}
        \end{equation}
        while
        \begin{equation}
            \vec{B}_{\mathrm{CD}}(\Bt) = \frac{(\BO \cdot \nabla_{\theta} \vec{B}) \times \vec{B}}{|\vec{B}|^2}
        \end{equation}
        is the counterdiabatic drive~\cite{Kolodrubetz2017}.

        The counterdiabatic drive is carefully chosen so that the quasienergy states of Eq.~\eqref{eqn:quasi_states} are created by
        \begin{align}
            c^\dagger_{x+} &= \frac{1}{\sqrt{2(1+\hat{B}_z)}}\left[ (1+\hat{B}_z) c^\dagger_{x \uparrow} + (\hat{B}_x + i \hat{B}_y) c^\dagger_{x\downarrow} \right], \nonumber \\
            c^\dagger_{x-} &= \frac{1}{\sqrt{2(1-\hat{B}_z)}}\left[ -(1-\hat{B}_z) c^\dagger_{x \uparrow} + (\hat{B}_x + i \hat{B}_y) c^\dagger_{x\downarrow} \right].
            \label{eqn:createpm}
        \end{align}
        Indeed, the counterdiabatic part \(\vec{B}_{\mathrm{CD}}\) is constructed to cancel the inertial term when moving to a frame co-rotating with \(\vec{B}\). The \(c^\dagger_{x\pm}\) operators have the property that
        \begin{equation}
            n_{x\pm} = c^\dagger_{x\pm} c_{x\pm} = c^\dagger_{x\mu} \left[ \tfrac{1}{2}(\mathbbm{1}\pm\hat{B}\cdot\vec{\sigma})\right]_{\mu\nu} c_{x\nu}
            \label{eqn:numpm}
        \end{equation}
        projects onto states with a fermion on site \(x\) with its spin aligned along \(\hat{B} = \vec{B}/|\vec{B}|\).

        The single-particle quasienergy states \(c^\dagger_{x\pm}(\Bt)\ket{0}\) (where \(\ket{0}\) is vacuum state) carry equal and opposite Chern numbers~\cite{Crowley2019}. That is, the Berry curvature
        \begin{equation}
            F = \nabla_\theta \times A,
            \quad\text{with}\quad
            A = i \bra{0} c_{x\pm} \nabla_{\theta} c^\dagger_{x\pm}\ket{0},
        \end{equation}
        has a nonzero quantized integral over the torus. (The Chern number is \(C = \pm 1\) when \(\vec{B}\) is given by Eq.~\eqref{eqn:B}.) The quantized average energy pumping between the drives is, in frequency lattice language, the quantized Hall current induced by the electric field \(\BO\) in the states \(c^\dagger_{x\pm}(\Bt)\ket{0}\). The \((\pm)\) modes each pump energy in a different direction---if \((+)\) pumps energy from drive 1 to 2, then \((-)\) pumps from drive 2 to 1 (\autoref{fig:Chernlayers}; cf. Ref.~\cite{Lapierre2021}).

        We refer to the modes \(c^\dagger_{x\pm}\) as \emph{pumping modes}.

    \subsection{Coupled layer model}
        \label{subsec:coupling}

        To complete the coupled layer construction we must add hopping terms between the sites.

        However, there is a complication because the pumping mode creation operators~\eqref{eqn:createpm} cannot be defined with a smooth gauge as a function of \(\Bt\). In Eq.~\eqref{eqn:createpm}, we have chosen a particular gauge with a phase singularity at the south pole of the Bloch sphere for \(c^\dagger_{x+}\), and at the north pole for \(c^\dagger_{x-}\).

        The number operators are gauge invariant, and so do not have this discontinuity, but a hopping term like \(c^\dagger_{x+} c_{x'-}\) will. Indeed,
        \begin{multline}
            c^\dagger_{x+} c_{x'-} = -\frac{1}{2} \sqrt{1-\hat{B}_z^2} \left[  c^\dagger_{x\uparrow} c_{x'\uparrow} - c^\dagger_{x\downarrow} c_{x'\downarrow} \right.\\
            \left.+ \frac{-\hat{B}_x + i \hat{B}_y}{1 - \hat{B}_z} c^\dagger_{x\uparrow} c_{x'\downarrow} + \frac{\hat{B}_x + i \hat{B}_y}{1 + \hat{B}_z} c^\dagger_{x\downarrow} c_{x'\uparrow} \right]
        \end{multline}
        has a phase singularity in the spin-flipping terms near \(\hat{B}_z = \pm 1\). This term cannot be included in the Hamiltonian if it is not a smooth quasiperiodic function. Fortunately, the norm of the hopping term need not be constant, so one can just arrange for the hopping term to vanish when it would otherwise have a singularity. The term
        \begin{equation}
            h_{x+,x'-} = \sqrt{1-\hat{B}_z^2}\; c^\dagger_{x+} c_{x'-}
        \end{equation}
        has no singularity, and is proportional to the desired hop.

        The full coupled layer Hamiltonian consists of three terms,
        \begin{equation}
            H(\Bt) = H_0 + H_{\mathrm{dis}} + H_{\mathrm{hop}}.
            \label{eqn:H_QP pump}
        \end{equation}
        The single-site part \(H_0\) is defined in Eq.~\eqref{eqn:H0}.

        The hopping term, written for open boundary conditions as
        \begin{multline}
            H_{\mathrm{hop}} = J\sum_{x=0}^{L-2} (1-\epsilon) h_{x+,x-} + \epsilon h_{(x+1)+,x-}  \\
            + J(1-\epsilon) h_{(L-1)+,(L-1)-} + \mathrm{H.c.},
            \label{eqn:H_hop}
        \end{multline}
        couples a (\(+\)) mode to a (\(-\)) mode, either within a site or between sites (\autoref{fig:wires}). The tuning parameter \(\epsilon\), which controls how large intersite hops are compared to intrasite hops, is the main variable of concern. All other parameters of the model will typically be fixed. \(H_{\mathrm{hop}}\) should be regarded as a Su-Schrieffer-Heeger (SSH) hopping term in a quasiperiodically rotating frame~\cite{Su1979}.

        Finally, the disorder term
        \begin{equation}
            H_{\mathrm{dis}} = \sum_{x = 0}^{L-1} \delta_{x+} n_{x+} + \delta_{x-} n_{x-}
            \label{eqn:H_dis}
        \end{equation}
        ensures the localization of fermions in the model~\cite{Anderson1958,Abrahams2010}. Each \(\delta_{x\pm}\) is taken to be uniformly random in \(\pm \Delta + [-\delta, \delta]\). We have included an on-site splitting of \(2 \Delta\) for greater control over the localization properties of the model (Appendix~\ref{app:frq_lat_loc}).

        Inspection of the limits \(\epsilon \in \{0,1\}\) reveals the properties of this model. When \(\epsilon = 0\), Hamiltonian~\eqref{eqn:H_QP pump} does not couple different sites and so is topologically trivial for any \(J \neq 0\) (\(W=0\)). On the other hand, when \(\epsilon = 1\) and with open boundary conditions, the edge modes \(n_{0+}\) and \(n_{L-}\) are uncoupled, and thus each pumps energy between the drives (\(W=1\) when \(\vec{B}\) is given by Eq.~\eqref{eqn:B}) (Figs.~\ref{fig:wires} and \ref{fig:Chernlayers}). Between these two limits, the edge states are deformed away from being perfectly localized to a single site, but cannot be destroyed unless the bulk delocalizes in either real space or in the synthetic dimensions, or both.

        With periodic boundary conditions, the ensemble of Hamiltonians also has a duality
        \begin{equation}
            \epsilon \mapsto 1-\epsilon, \quad
            c_{x +} \mapsto c_{x -}, \quad
            c_{x -} \mapsto c_{(x+1) +},
        \end{equation}
        which maps topological phases to trivial phases, and vice versa. Thus, if there is a unique critical point between these phases, it must be at the self-dual point \(\epsilon = \epsilon_c = 1/2\).

        We note that the coupled layer construction also makes the bulk circulation of energy in the QP pump intuitive (\autoref{subsec:QP-pump}). When pumping modes between different sites are coupled, they (at a cartoon level) hybridize into small circulating loops (\autoref{fig:wires}). This is the bulk energy circulation.

\section{Numerical Characterization}
    \label{sec:numerics}

    The edge states of model~\eqref{eqn:H_QP pump} can be found exactly when \(\epsilon = 1\). They are created by \(c^\dagger_{0 +}\) and \(c^\dagger_{(L-1) -}\). At \(\epsilon = 0\) all couplings are intrasite, and the phase is trivial. At \(\epsilon=1/2\) the model is self-dual, and cannot be localized. Away from these limits, we resort to numerics to find properties of the coupled layer model~\eqref{eqn:H_QP pump}.

    The steady states of localized quasiperiodically driven models may be extracted through exact diagonalization of the frequency lattice quasienergy operator~\eqref{eqn:frq_lat}. This method is resource-intensive. It requires expanding the Hilbert space with the auxiliary drive states \(\ket{\Bn}\) and truncating the frequency lattice Hilbert space. We will instead focus on observables that can be measured from real time dynamics using a numerical solution of the Schr\"odinger equation (for which we use the ordinary differential equation methods of \textsc{quspin}~\cite{Weinberg2017,Weinberg2019}), namely, the lattice site occupation numbers and the energy transferred between the drives (more precisely, the work done on the system by the drives~\eqref{eqn:pow}).

    Our numerics recover expected properties of the topological and trivial phases of the QP pump, including localization and the pumping of energy at the edge (\autoref{subsec:loc_phase}). Finite-size scaling analysis of the energy transferred between the drives finds a scaling collapse consistent with a single critical point (\autoref{subsec:crit_point}). The phenomenology and critical exponents of the transition suggest it lies in the universality class of the two-dimensional quantum Hall transition.

    \subsection{Phases}
        \label{subsec:loc_phase}

        Localization is a necessary ingredient of the QP pump. To probe this numerically, we compute the lattice site occupation numbers of an initially localized fermion:
        \begin{equation}
            n_x(t) = \qexp{(c^\dagger_{x \uparrow}c_{x \uparrow} + c^\dagger_{x \downarrow} c_{x \downarrow})}{\psi_0(t)}.
        \end{equation}
        Here, \(\ket{\psi_0(0)} = c^\dagger_{0 \uparrow}\ket{0}\) is the initial state with one spin up fermion at \(x=0\), \(\ket{\psi_0(t)}\) is the corresponding time evolved state under Hamiltonian~\eqref{eqn:H_QP pump}, and \(c^{(\dagger)}_{x \mu}\) is a fermion annihilation (creation) operator at site \(x\) and with spin \(\mu\). Calculations of \(n_x(t)\) are performed with periodic boundary conditions to avoid the effects of the pumping edge modes.

        The typical late time value of \(n_x(t)\) is computed as
        \begin{equation}
            \ln [n_x]_{\mathrm{typ}} = \left[\frac{2}{T}\int^{T}_{T/2} \d t \ln n_x(t)\right]_{\mathrm{mean}},
            \label{eqn:ntyp}
        \end{equation}
        where \([\cdot]_{\mathrm{mean}}\) denotes an average over the disorder realization \(\{\delta_{x\pm}\}\) and initial phase \(\Bt_0\), and \(T\) is a time much larger than all inverse energy scales in the problem. We use the typical value (geometric mean) for \(n_x\) as a forward scattering approximation predicts that \(n_x\) is log-normally distributed across disorder realizations for fixed \(x\)~\cite{Abrahams2010}. This makes the typical value a more meaningful estimate for the center of the distribution.

        \begin{figure}
            \centering
            \includegraphics{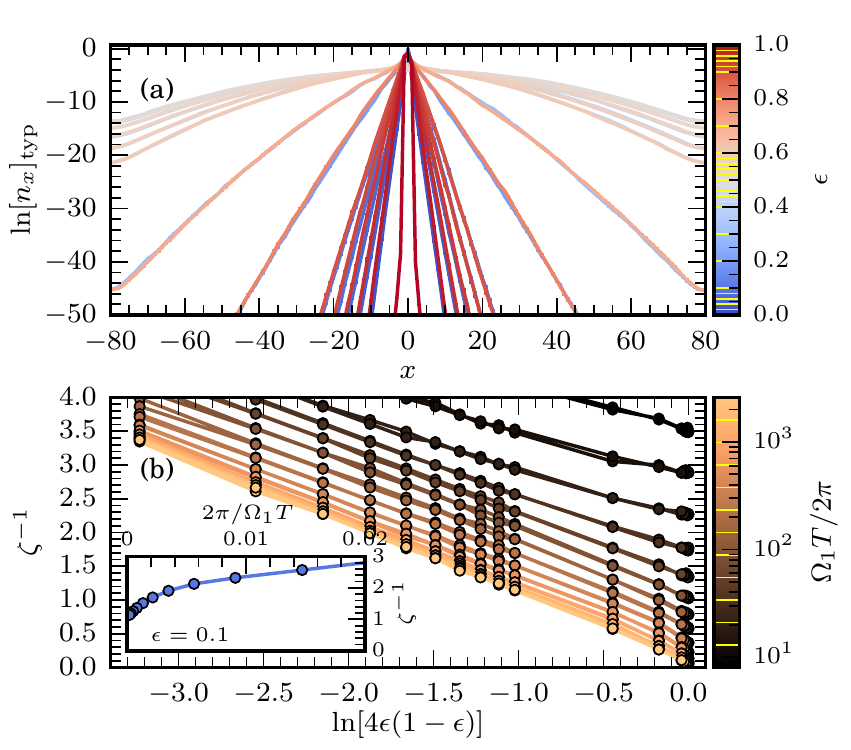}
            \caption{(\textbf{a}) The typical occupation \(\ln [n_x]_{\mathrm{typ}}\) decays for different \(\epsilon\). There is clear exponential decay for small \(\epsilon(1-\epsilon)\). When \(\epsilon \approx 1/2\) is close to the critical value, the late time \(\ln [n_x]_{\mathrm{typ}}\) has not yet converged, and appears parabolic in \(x\), indicative of diffusive dynamics. (\textbf{b}) Fitting the localization length from the decay of \(\ln [n_x]_{\mathrm{typ}} \sim -2x/\zeta\) shows the expected \(1/\ln[\epsilon(1-\epsilon)]\) scaling. \textbf{Inset}: For small \(\epsilon (1-\epsilon)\), \(\zeta^{-1}\) converges to a nonzero value as \(T\) is increased. \emph{Parameters:} \(L=160\) with periodic boundaries, \(B_0/\Omega_1 = 1\), \(\Omega_1/\Omega_2 = (1+\sqrt{5})/2\), \(\Omega_1 T/(2\pi) = 2584\), \(\delta/\Omega_1 = 0.09\), \(\Delta/\Omega_1 = 0.7\), and \(J/\Omega_1 = 0.305\). 300 disorder configurations and initial phases are used for averages in \(\ln [n_x]_{\mathrm{typ}}\), with plotted error bars giving one standard error of the mean (often too small to be visible). Values of \(\epsilon\) and \(T\) used are marked in yellow in the color bars.}
            \label{fig:corr_len}
        \end{figure}

        The occupation \(\ln [n_x]_{\mathrm{typ}}\) is plotted for several different values of \(\epsilon\) in \autoref{fig:corr_len}(a). 

        Many features of \([n_x]_{\mathrm{typ}}\) follow from the coupled layer construction, or standard results in the theory of Anderson localization~\cite{Anderson1958,Abrahams2010}. When \(\epsilon \in \{0,1\}\), the model is perfectly localized---the occupations \(n_x(t)\) can only be nonzero for \(x=0\) in the trivial phase, or \(x \in \{0,\pm 1\}\) in the topological phase.

        As \(\epsilon\) is moved away from these limits, \([n_x]_{\mathrm{typ}}\) remains exponentially decaying in \(|x|\), but the localization length \(\zeta\) increases. Standard estimates from Anderson localization give that \(\zeta^{-1} = O(\ln t/\delta)\), where \(t\) is a hopping amplitude, \(\delta\) is the disorder strength, and \(t \ll \delta\). The coupled layer model has modulated strong and weak hops between pumping modes, so it is more meaningful to use an amplitude associated to double hops spanning both a weak and strong bond---from \(n_{x+}\) to \(n_{(x+1)+}\). Second order perturbation theory predicts that this effective hopping is proportional to \(t = J^2\epsilon(1-\epsilon)/\Delta\). Thus
        \begin{equation}
            \zeta^{-1} = O\left(\ln\left[ \frac{J^2}{\Delta \delta} \epsilon(1-\epsilon) \right]\right)
            \quad\text{for}\quad
            \frac{J^2}{\Delta}\epsilon(1-\epsilon) \ll \delta.
            \label{eqn:zeta_pred}
        \end{equation}
        We see this scaling in \autoref{fig:corr_len}(b).

        \begin{figure}
            \centering
            \includegraphics{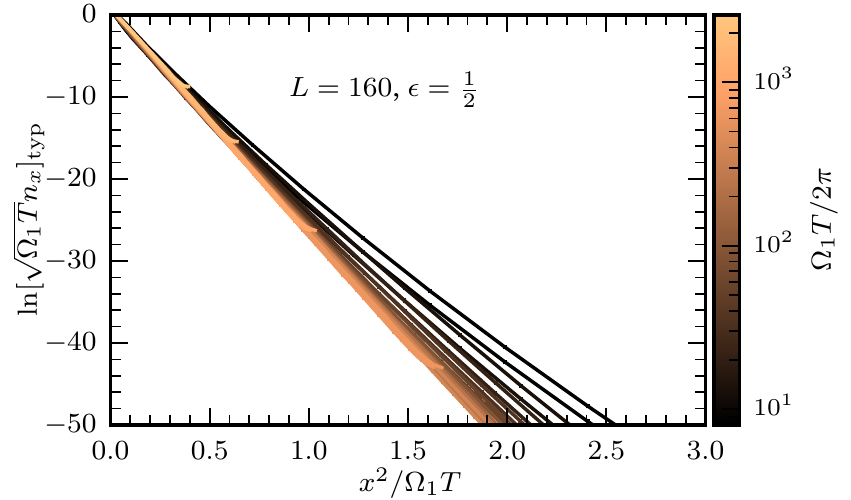}
            \caption{Rescaling \(x^2\) by time (\(T\), the integration time~\eqref{eqn:ntyp}) produces data collapse in \(\ln[n_x]_{\mathrm{typ}}\) at \(\epsilon=1/2\), consistent with diffusive dynamics~\eqref{eqn:nx_z2}. The collapse is improved by including the subleading correction predicted by diffusion, \(\tfrac{1}{2}\ln \Omega_1 T\).  \emph{Parameters:} \(\epsilon = 1/2\), and as in \autoref{fig:corr_len}.}
            \label{fig:rescale_ntyp}
        \end{figure}

        Close to the self-dual point \(\epsilon = 1/2\), \(\ln [n_x(t)]_{\mathrm{typ}}\) appears parabolic at numerical time scales:
        \begin{equation}
            \ln [n_x(t)]_{\mathrm{typ}} \sim -\frac{x^2}{D t},
            \label{eqn:nx_z2}
        \end{equation}
        with some \(D>0\). Indeed, rescaling \(x^2\) by \(t\) produces data collapse in \(\ln [n_x(t)]_{\mathrm{typ}}\) for small \(x^2/t\) (\autoref{fig:rescale_ntyp}).

        Equation~\eqref{eqn:nx_z2} is characteristic of a diffusive regime in dynamics. The finite-size scaling analysis of \autoref{subsec:crit_point} suggests that for \(\epsilon \neq 1/2\), this diffusive behavior is a finite-size effect associated to an isolated critical point, rather than a diffusive phase.

        While localization is vital for the stability of the QP pump, it does not reveal its topological properties. The energy transferred between the drives is the interesting observable in this context, and it is this we use to numerically demonstrate the presence of the topological phase. Specifically, the order parameter in \autoref{fig:pump_PD} is the average rate of energy transfer between the drives, \(\bar{P}\)~\eqref{eqn:Pbar}.

        \begin{figure}
            \centering
            \includegraphics{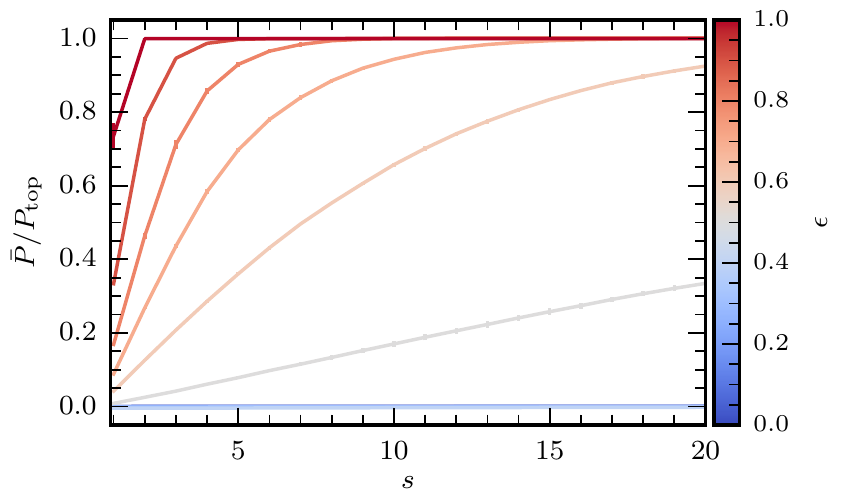}
            \caption{The topological edge modes responsible for pumping are exponentially localized. This is revealed by computing the dependence of the average pumping rate \(\bar{P}\) on the filling \(s\) (\autoref{fig:pump_PD}). \(\bar{P}\) converges exponentially to its topological value \(\Ptop W\) outside a critical region around \(\epsilon = 1/2\). \emph{Parameters:} As in \autoref{fig:corr_len}, but with open boundary conditions. \(\bar{P}\) is averaged over 200 disorder and initial phase samples. \(\epsilon \in [0,1]\) is taken in steps of \(0.1\). (Note that all curves with \(\epsilon \leq 0.4\) overlap.)}
            \label{fig:edge_modes}
        \end{figure}

        To numerically measure \(\bar{P}\), we integrate the expectation value of \(P(t) = -\Omega_2 \partial_{\theta_2} H\) (Eq.~\eqref{eqn:pow}), which gives the power transferred into the drive of frequency \(\Omega_2\). The total work done on this drive is
        \begin{equation}
            \Delta E_2(t) = \int_0^t \d t'\, \tr{P(t')\rho(t')},
        \end{equation}
        where \(\rho(t')\) is a time evolved state.

        The initial state \(\rho(0)\) would, ideally, be the pumping mode itself. This is difficult to prepare, and even numerically we do not know its precise form. However, as the pumping mode is localized near the edge, taking \(\rho(0) = \rho_s\), the Slater-determinant state with the first \(s\) sites near the edge filled (\autoref{fig:pump_PD}(a)), ensures the pumping mode is completely occupied, up to an exponentially small weight outside the range \(s\) (\autoref{fig:edge_modes}). No other modes pump, except the edge mode at the opposite edge, so all pumping is due to the occupied edge mode. Thus, one expects to find
        \begin{equation}
            \Delta E_2(t) = \Ptop W t + O(e^{-s/\zeta}, t^0)
            \label{eqn:DE_Ptop}
        \end{equation}
        when \(\rho=\rho_s\). (In fact, one may also populate any additional sites in the bulk much further than \(\zeta\) from the edges. This does not affect the average pumping rate, but our numerics do not use such states.)

        The late time average \(\Delta E_2 / t\) converges to \(\bar{P}\), but several numerical techniques can make the estimation of \(\bar{P}\) more reliable. While Eq.~\eqref{eqn:DE_Ptop} holds in each disorder realization and for any initial phase \(\Bt_0\), averaging \(\Delta E_2(t)\) over disorder and initial phase reduces the subleading corrections for finite \(s\) and \(t\). Then, fitting the late time data (we use the last half of the observed time series) to a straight line provides an estimate for the average pumping rate \(\bar{P}\) which biases the longest time scales.

        In \autoref{fig:pump_PD}(b), we find that \(\bar{P}\) is quantized to the expected topological values of \(0\) (small \(\epsilon\)) or \(1\) (large \(\epsilon\)) outside of a critical region near the self-dual point (\(\epsilon = 1/2\)). Further, this critical region sharpens with increasing \(L\) and \(s\), suggesting the smooth crossover could be a finite size effect. The exponential convergence of \(\bar{P}\) to the quantized value with increasing \(s\) is shown in \autoref{fig:edge_modes}.

    \subsection{Critical point}
        \label{subsec:crit_point}

        In previous numerical studies of the QP pump, it has not been clear whether the topological and trivial phases are separated by an isolated critical point, or an intervening critical phase~\cite{Nathan2020b,Long2021}. The coupled layer model enjoys a self-duality which fixes a value that must be delocalized, \(\epsilon = 1/2\), and simplifies finite-size scaling analysis. Our findings are consistent with \(\epsilon = 1/2\) being an isolated critical point.

        The dynamical exponent, \(z\), describes the scaling relationship between length and time at the critical point. Prompted by the parabolic shape of late time \(\ln[n_x(t)]_{\mathrm{typ}}\), it is natural to suspect that the critical point is diffusive, with \(z = 2\)~\cite{Hohenberg1977}.

        If, indeed, \(z=2\), rescaling time as \(t/x^2\) (with \(x\) some length scale) should produce data collapse in observable quantities. The length scale we rescale by is \(s\), the finite extent of the initial Slater-determinant state. The observable we inspect is \(\Delta E_2(t, s, \epsilon)\) (noting the dependence on \(s\) and \(\epsilon\) explicitly). (Rescaling \(x^2\) by \(t\) in \(\ln[n_x(t)]\) also produces data collapse as in Eq.~\eqref{eqn:nx_z2}; see \autoref{fig:rescale_ntyp})

\begin{figure}
            \centering
            \includegraphics[width=\linewidth]{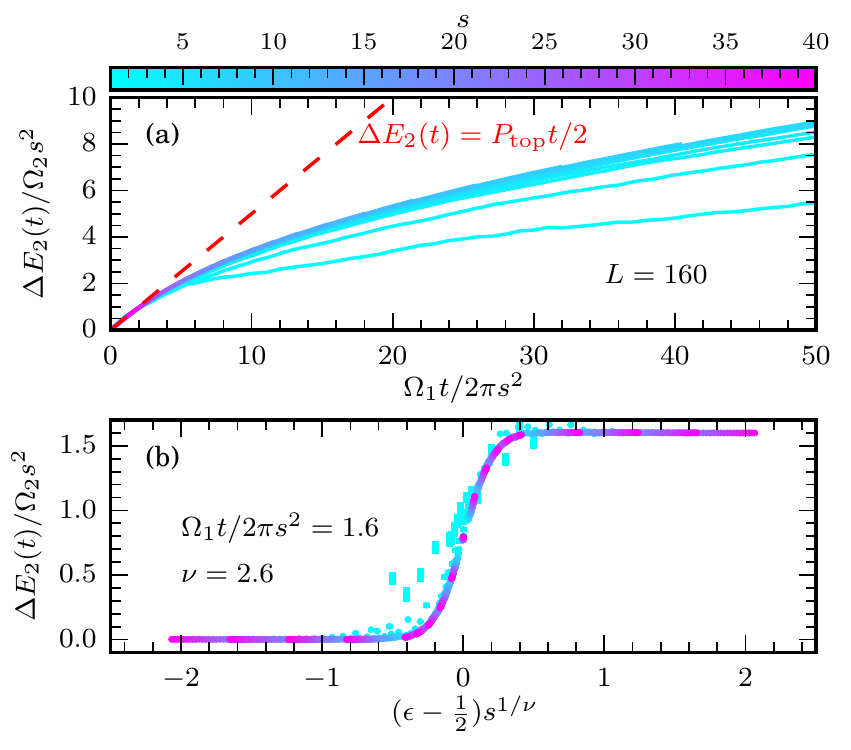}
            \caption{Finite-size scaling collapse around the critical point. (\textbf{a}) Rescaling \(t\) by \(s^2\) at the critical point \(\epsilon=\epsilon_c = 1/2\) collapses the energy curves \(\Delta E_2(t)\), showing that the critical point is diffusive (dynamical exponent \(z = 2\)). At short times, \(\Omega_1 t/2\pi s^2 \ll 1\), the average power is half the topological value. (\textbf{b}) Rescaling \(\epsilon-1/2\) by \(s^{1/2.6}\) produces a good data collapse for \(E_2\) at fixed \(\Omega_1 t/2\pi s^2\) and large \(s\), consistent with the critical exponent \(\nu\) for the two-dimensional quantum Hall effect transition. \emph{Parameters:} As in \autoref{fig:corr_len} with open boundary conditions. \(\Delta E_2(t)\) is averaged over 200 disorder and initial phase samples.}
            \label{fig:fss}
        \end{figure}

        \autoref{fig:fss}(a) shows that the measured \(\Delta E_2(t, s, \epsilon=1/2)\) is consistent with the scaling form
        \begin{equation}
            \Delta E_2(t, s, \epsilon=1/2) \sim s^z \mathcal{E}_2(t/s^z)
        \end{equation}
        with \(z = 2\). With this scaling relation, the lines \(\Delta E_2 \propto t\) are fixed. Additionally, well before the time scale for diffusion by length \(s\), \(\Omega_1 t /s^2 \ll 1\)~\footnote{In identifying \(s^2/ \Omega_1\) as the timescale for diffusion by length \(s\), we assume that the diffusion constant is set by dimensional analysis: \(D = O(a^2 \Omega_1)\), where \(a\) is the lattice constant. \autoref{fig:fss}(a) indicates \(\Omega_1 t /s^2 \ll 1\) is sufficient to see half-quantized pumping, from which we infer that diffusion away from the edge is limited on this timescale.}, the rate of pumping is precisely half the quantized value:
        \begin{equation}
            \Delta E_2(t,s,\epsilon=1/2) \sim \Ptop t/2, \quad \Omega_1 t/s^2 \ll 1.
        \end{equation}
        By making \(s\) larger, this half-integer pumping can be made to persist for an arbitrarily long time.

        If the \(\epsilon=1/2\) critical point is isolated, then a nonzero value of \(\epsilon-1/2\) introduces a finite localization length \(\zeta\). The divergence of \(\zeta\) defines another important critical exponent, \(\nu\):
        \begin{equation}
            \zeta \sim A (\epsilon-\tfrac{1}{2})^{-\nu}.
        \end{equation}
        We assess a corresponding scaling form for \(\Delta E_2(t,s,\epsilon)\),
        \begin{equation}
            \Delta E_2(t, s, \epsilon) \sim s^z \mathcal{E}_2(t/s^z, (\epsilon-\tfrac{1}{2})s^{1/\nu}).
            \label{eqn:rescale_nu}
        \end{equation}

        There is a relatively broad range of \(\nu \in [2.2, 2.8]\) which produce acceptable collapse in our data. The particular value \(\nu = 2.6\) is shown in \autoref{fig:fss}(b). Beyond some small \(s\) transient behavior, all rescaled data lie on the same curve.

        The scaling form~\eqref{eqn:rescale_nu} with any positive \(\nu\) suggests that keeping \(s/L\) fixed and taking \(s \to \infty\) faster than \(\sqrt{t}\) sharpens the curve for \(\bar{P}\) in \autoref{fig:pump_PD} to a step function. (In \autoref{fig:pump_PD}, \(t=T\) is taken to be a large fixed value.) Further, there is a unique critical point with diffusive dynamics and a half-integer energy current, \(\bar{P} = \Ptop /2\).

        We note that half-integer quantization of the topological response has long been recognized in the context of the integer quantum Hall effect~\cite{Khmelnitskii1983,Levine1983,Laughlin1984,Lee1992,Kivelson1992}, and more recently in the analogous setting of quasiperiodically driven spins~\cite{Crowley2020,Boyers2020}. In the quantum Hall context, the analogous quantity to the pumping rate is the Hall conductivity \(\sigma_{xy}\), the scaling theory for which predicts an unstable half-integer fixed point at the transition~\cite{Khmelnitskii1983,Levine1983,Laughlin1984,Lee1992,Kivelson1992}. For noninteracting particles, the critical point is also diffusive (\(z=2\)), with a critical exponent for the divergence of the correlation length \(\nu = 2.593 \pm 0.005\)~\cite{Chalker1988,Slevin2009,Puschmann2019}.

        The quantum Hall phenomenology is consistent with our observations in \autoref{fig:fss}. It is tantalizing to make strong comparisons between the integer quantum Hall transition and the QP pump transition, especially given the cartoons for the QP pump in Figs. \ref{fig:wires} and \ref{fig:Chernlayers}.

        Nonetheless, a precise argument indicating that the QP pump transition is in the universality class of the two-dimensional integer quantum Hall transition remains elusive. Quantum Hall systems and the QP pump share a similar coupled layer construction (\autoref{fig:wires}), but the disorder along \(\hat{\Omega}_\perp\) in the QP pump is correlated~\cite{Long2022a}. This may alter critical exponents compared to those with uncorrelated disorder.

        Further, comparing \(\bar{P}\) and \(\sigma_{xy}\) in the picture of \autoref{fig:wires} is problematic---\(\sigma_{xy}\) is defined in terms of linear response, while \(\bar{P}\) arises from edge physics in the coupled layer picture. Especially at the transition, it becomes unclear why a half-integer \(\sigma_{xy}\) should relate to a half-integer \(\bar{P}\). Recent work has explored the presence and nature of edge states at the transitions between topological phases, including in quantum Hall systems~\cite{Verresen2020,Balabanov2022,Fu2022}. However, a complete understanding of the edge state properties has not yet been achieved.

\section{Effect of Interactions}
    \label{sec:interactions}

    The QP pump is proposed to be an infinitely long lived phase of matter even with weak interactions~\cite{Long2021,Long2022a}. Localization is essential here, as it protects the system from absorbing energy from the drives and heating to a featureless infinite temperature state~\cite{Basko2006,Oganesyan2007,Huse2013}. While the asymptotic stability of localization in interacting strongly disordered systems has recently been brought into debate~\cite{Abanin2021,Suntajs2020,Schulz2020,Taylor2021,Sels2021,Sels2021b,Morningstar2022,Sels2021a,Sierant2022}, it remains universally accepted that such systems remain localized for a sufficiently long time to give rise to prominent \emph{prethermal} regimes~\cite{Mori2018}. The existence of such a prethermal regime also extends to quasiperiodically driven systems~\cite{Else2019}.

\begin{figure}
        \centering
        \includegraphics{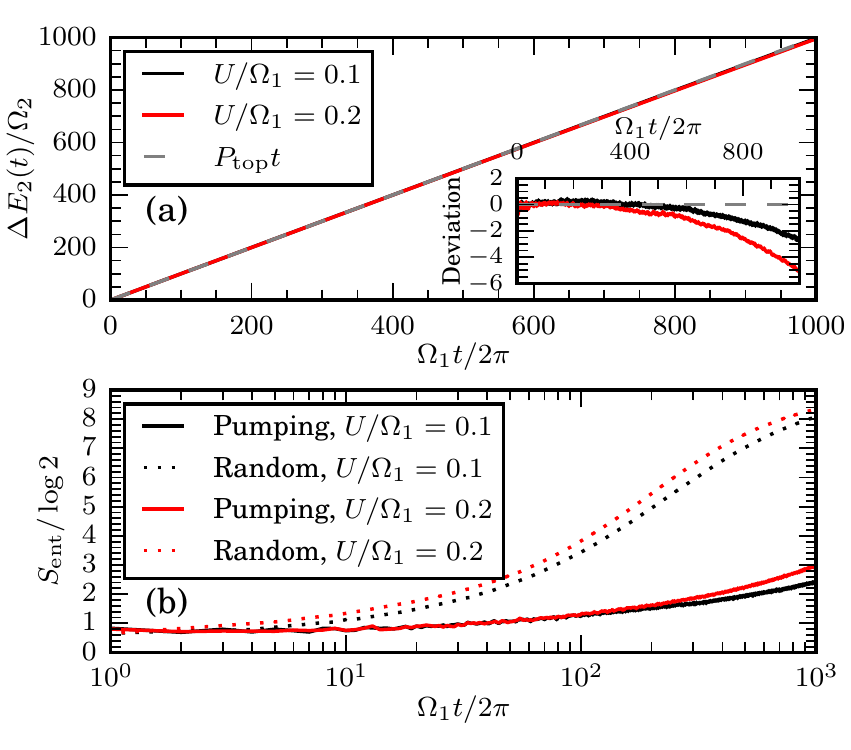}
        \caption{(\textbf{a}) The energy pumped into drive 2 is very close to the topological value \(\Delta E_2(t) = \Ptop t\), even with nonzero interaction strength \(U\). This indicates the existence of a long-lived prethermal regime where topological pumping persists. (\textbf{b}) Nonetheless, these parameter values are delocalized. The half-cut entanglement entropy, \(S_{\mathrm{ent}}\), increases faster than logarithmically to its saturation value in a random initial product state, indicative of thermalization. The pumping state with \(s\) sites filled from the \(x=0\) edge, which is far from random, thermalizes much slower. \emph{Parameters:} \(L = 10\), \(s = 5\), \(\Omega_1 T/2\pi = 1000\), \(\epsilon = 0.9\). \(\Delta E_2(t)\) and \(S_{\mathrm{ent}}\) are averaged over \(200\) samples of initial product states, disorder realizations and initial phases. All other parameters as in \autoref{fig:corr_len}.}
        \label{fig:interacting_pump}
    \end{figure}

    Topological pumping persists when adding weak interactions to the coupled layer model (\autoref{fig:interacting_pump}). In the parameter regimes accessible by our numerics, this behavior is prethermal. It persists for a long, but finite, time in any finite size system.

    We consider the time-dependent interaction (recall that the pumping mode number operators \(n_{x\pm}(\Bt)\) depend quasiperiodically on time, Eq.~\eqref{eqn:numpm})
    \begin{equation}
        H_{\mathrm{int}}(\Bt) = U \sum_{x = 0}^{L-1}  \left[n_{x+} n_{x-} + n_{x-} n_{(x+1)+}\right],
    \end{equation}
    which preserves the self-duality of model~\eqref{eqn:H_QP pump}. This ensures that any localized trivial phase with \(\epsilon < 1/2\) must be mirrored by a localized topological phase with \(\epsilon > 1/2\).

    \autoref{fig:interacting_pump}(a) shows that the transferred energy \(\Delta E_2 (t)\) in the interacting model \(H + H_{\mathrm{int}}\) is extremely close to the noninteracting prediction. Even after \(1000\) cycles of the first drive, only around \(5\) of the expected \(1000\) energy quanta have not been pumped into drive 2.

    However, the model is delocalized for the parameters in \autoref{fig:interacting_pump}. Localization in the interacting model can be assessed by measuring the half-cut entanglement entropy, \(S_{\mathrm{ent}}\), for an initial product state. In a localized phase, \(S_{\mathrm{ent}}\) should increase logarithmically until it eventually saturates due to finite-size effects~\cite{Bardarson2012,Serbyn2013,Huse2014}. \autoref{fig:interacting_pump}(b) shows that \(S_{\mathrm{ent}}\) appears to increase faster than logarithmically prior to finite-size saturation, indicating that the system is not localized in accessible parameter regimes.

    This lack of localization can be understood through the analysis of Ref.~\cite{Long2022a}. Reference~\cite{Long2022a} finds that the QP pump is stable, even with interactions, provided that the many-body localization length \(\xi\) is below a critical value~\footnote{Here, we are taking the local Hilbert space dimension to be \(4\), as is appropriate for spinful fermions, unlike Ref.~\cite{Long2022a}, which considered qubit chains.}
    \begin{equation}
        \xi \leq \xi_c = (2 \ln 4)^{-1}.
        \label{eqn:xi_crit}
    \end{equation}
    The noninteracting coupled layer model is arbitrarily well-localized for \(\epsilon(1-\epsilon) \ll 1\), but Eq.~\eqref{eqn:zeta_pred} and \autoref{fig:corr_len} show that the single-particle localization length only approaches zero logarithmically. Comparison of \autoref{fig:corr_len} to Eq.~\eqref{eqn:xi_crit} shows that \(\epsilon \lesssim 0.05\) (or \(1-\epsilon \lesssim 0.05\)) is necessary to have \(\zeta < \xi_c\). Further, even \(\epsilon \lesssim 0.05\) is likely an overestimate for the stability of localization. Interactions should be expected to renormalize \(\zeta\) significantly when calculating the many-body localization length \(\xi\). Thus, very small values of \(\epsilon\) would be required to observe asymptotic many-body localization (MBL) in the coupled layer model. This is problematic in finite time numerics, as an integration time of many times \(2\pi/\epsilon\) is required to observe the effects of the hopping term, and thus even have the possibility of observing thermalization.

    Part of the reason for the extremely long lifetime of pumping is the highly nonthermal initial state, \(\rho_s\), in which the system is prepared. It takes much longer for the fermions to diffuse from the left-hand side (say) of the system to uniformity than it does for a random initial distribution to thermalize. This can be seen by comparing the half-cut entanglement entropy of the pumping initial state to a random product state (\autoref{fig:interacting_pump}(b)). The entropy in the random state increases faster than logarithmically to its saturation value, while the pumping state entropy increases very little on the observed timescale.

\section{General Construction}
    \label{sec:general_model}

    Any \((1+D)\)-dimensional ALTP can be constructed through coupled layers, as in \autoref{sec:model}. With a careful coupling of critically tuned sites driven by \(D\) incommensurate frequencies, these models retain the self-duality properties from the \(D=2\) case.

    The uncoupled starting point of the construction is formally similar to \autoref{sec:model}:
    \begin{equation}
        H_0(\Bt) = \sum_{x=0}^{L-1} c^\dagger_{x \mu}\left[-(\vec{B}+\vec{B}_{\mathrm{CD}})\cdot\vec{\Gamma}\right]_{\mu \nu} c_{x \nu}.
        \label{eqn:H0_moretones}
    \end{equation}
    In the new context, \(\Bt\) is a vector of \(D\) drive phases, with the corresponding vector of frequencies \(\BO\); the \(\Gamma_j\) give some convenient operator basis for the single-site Hamiltonian; and \(B_j\) and \(B_{\mathrm{CD},j}\) are their coefficients. The eigenvalues of \(-\vec{B}(\Bt)\cdot\vec{\Gamma}\) form continuous \emph{bands} as a function of \(\Bt\). This structure is analogous to the band theory of solids, where \(\vec{k}\) (the crystal momentum) plays the same role as \(\Bt\).

    The uncoupled model~\eqref{eqn:H0_moretones} can be fine tuned to possess chiral topological states. When the number of tones \(D = 2n\) is even, \(\vec{B}(\Bt)\) can be chosen so that the projector \(p_-\) onto the eigenstates of \(-\vec{B}(\Bt)\cdot\vec{\Gamma}\) with negative energy---the \emph{lower bands}---has a nonzero \(n\)th Chern number, \(C_n\)~\cite{Teo2010}. (We will give a particular example for the second Chern number below.)  Then, \(\vec{B}_{\mathrm{CD}}(\Bt)\) should be chosen to eliminate excitations between the lower and upper bands induced by the drive. Writing \(p_-\) for the projector onto the lower bands, the necessary and sufficient condition for the suppression of excitations out of the lower bands is~\cite{Kolodrubetz2017}
    \begin{equation}
        [-i \partial_t B_j \Gamma_j + [B_k \Gamma_k, B_{\mathrm{CD},l} \Gamma_l ], p_-] = 0,
    \end{equation}
    (where summation is implied). This gives a linear equation for \(B_{\mathrm{CD},l}\) in terms of \(B_j(\Bt)\) and the coefficients \(f_{jkl}\) in an expansion of the commutator \([\Gamma_k, \Gamma_l] = f_{jkl}\Gamma_j\).
    \begin{equation}
        (i \BO \cdot \nabla_{\theta} B_j + B_k B_{\mathrm{CD},l} f_{jkl})[\Gamma_j, p_-] = 0
    \end{equation}
    Solutions to this equation are not unique.

    Now the sites must be coupled. A generic hopping between sites will typically allow for a localized phase, but to unambiguously identify the edge state in some limit the coupling must be carefully chosen. We denote the \(\Bt\)-dependent fermion annihilation operators in each band as \(c^{\pm}_{x \mu}(\Bt)\) with a superscript \(\pm\) depending on whether the band has positive or negative energy in \(-\vec{B}\cdot\vec{\Gamma}\). The definition of these operators require a choice of gauge, which cannot be smooth if any Chern number is nonzero. The hopping term between sites is, with open boundary conditions,
    \begin{multline}
        H_{\mathrm{hop}} = J_{\mu\nu}(\Bt) \sum_{x=0}^{L-2} (1-\epsilon) c^{+ \dagger}_{x \mu} c^{-}_{x \nu} + \epsilon c^{+ \dagger}_{(x+1) \mu} c^{-}_{x \nu}  \\
        + J_{\mu\nu}(\Bt) (1-\epsilon) c^{+ \dagger}_{(L-1) \mu} c^{-}_{(L-1) \nu} + \mathrm{H.c.}
    \end{multline}
    As the \(c^{+ \dagger}_{x \mu} c^{-}_{x' \nu}\) hopping terms are not smooth, the hopping coefficients \(J_{\mu\nu}(\Bt)\) must be chosen so as to vanish sufficiently quickly around any singularities, leaving \(H_{\mathrm{hop}}\) smooth and well-defined. Otherwise, there is significant freedom in the choice of \(J_{\mu\nu}(\Bt)\). Any choice leaves the upper bands (those with positive energy) uncoupled at \(x=0\) when \(\epsilon = 1\) (\autoref{fig:wires}).

    An on-site disorder term is responsible for localization:
    \begin{equation}
        H_{\mathrm{dis}} = \sum_{x = 0}^{L-1} \delta_{x+} p_{x+} + \delta_{x-} p_{x-},
    \end{equation}
    where \(p_{x\pm}\) is the projector onto the upper (lower) bands on site \(x\), and \(\delta_{x\pm}\) are independent and identically distributed random numbers.

    The total Hamiltonian is
    \begin{equation}
        H = H_0 + H_{\mathrm{hop}} + H_{\mathrm{dis}}.
    \end{equation}
    This model has the same \(\epsilon \leftrightarrow 1-\epsilon\) self-duality as the two-tone model. It has uncoupled, perfectly localized edge modes with \(n\)th Chern number \(\pm C_n\) when \(\epsilon =1\).

    The winding number invariant, \(W\), of the bulk model is given by \(C_n\) in the nontrivial phase.

    \subsection{Four-dimensional quantum Hall edge states}
        \label{subsec:4DQHE}

\begin{figure}
            \centering
            \includegraphics[width=\linewidth]{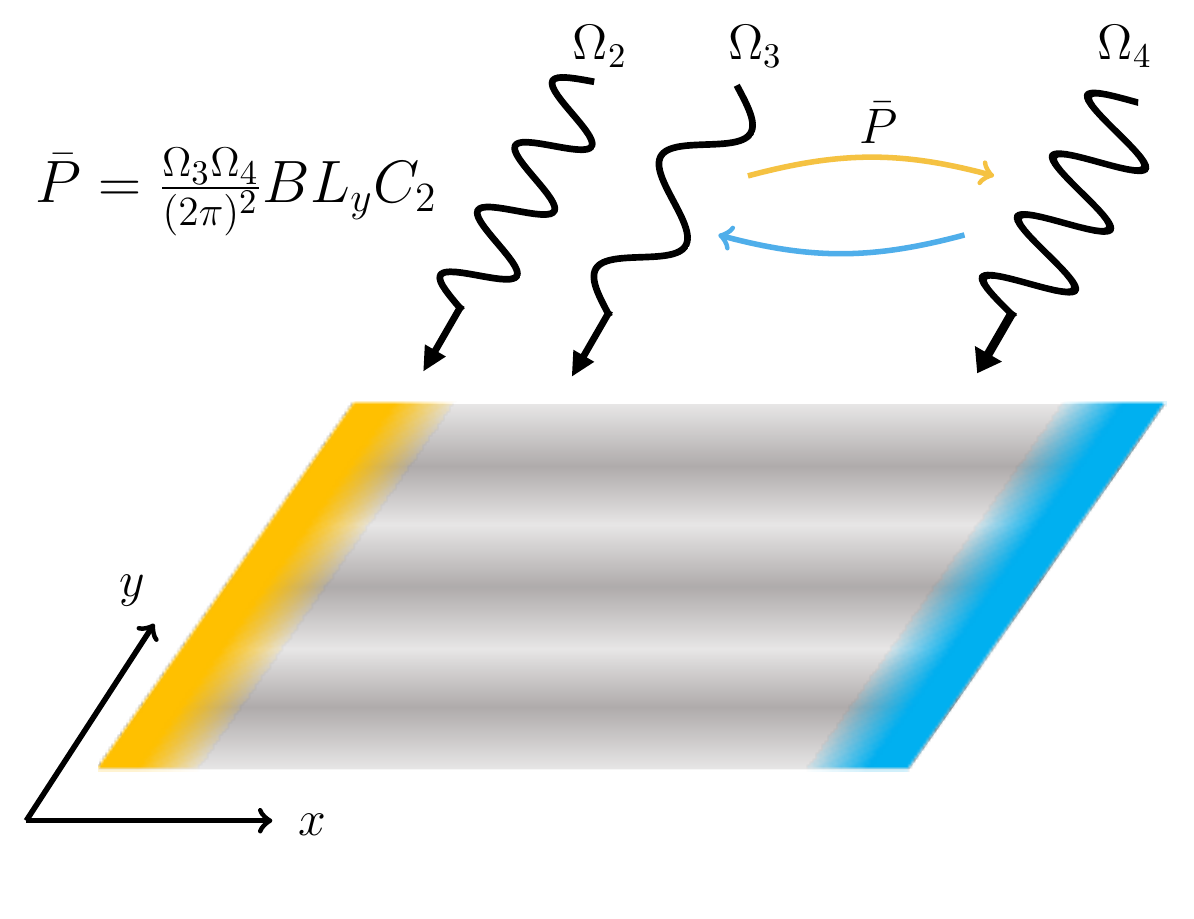}
            \caption{\label{fig:4DQHE}The coupled layer construction may be used to model a \((2+3)\)-dimensional ALTP. The edge states (yellow, blue) possess a nontrivial second Chern number \(\pm C_2\) which is equal (up to a sign depending on which edge is being considered) to the bulk winding number \(W\). Their response is analogous to the nonlinear response in the four-dimensional quantum Hall effect. A magnetic flux \(B\) per plaquette is introduced in the \((\hat{y},\hat{e}_2)\) plane by a linear winding of the initial phase \(\theta_{02} = B y\) in space (grey stripes). The average energy pumping rate between drives 3 and 4 depends on both the magnetic field \(B\) and the electric field \(\BO\): \(\bar{P}/L_y = \Omega_3 \Omega_4 B C_2/(2\pi)^2\). \(L_y\) is the length of the pumping edge.}
        \end{figure}

        The simplest ALTP beyond the QP pump has \(d+D = 5\). The general coupled layer construction shows that this phase has edges states with a nontrivial second Chern number \(C_2\), as appearing in the four-dimensional integer quantum Hall effect~\cite{Teo2010,Lohse2018}. In this section, we explore this case in more detail. In a related \((2+3)\)-dimensional model constructed from the coupled layer approach (\autoref{fig:4DQHE}), we describe the physical observable associated to the edge states---a nonlinear (in synthetic field strength) energy pumping response.

        Just as a nonzero first Chern number implies a quantized linear response to a weak electric field, a nonzero second Chern number implies a quantized quadratic response to an electric \emph{and} magnetic field. In the frequency lattice, the electric field is the vector \(\BO\). It is not possible to implement a magnetic field in the \((0+D)\)-dimensional geometry of the edge in the coupled layer construction. Instead, one should seek a \((2+3)\)-dimensional model (with a \((1+3)\)-dimensional edge), where a magnetic field in the synthetic dimensions may be emulated through a spatially dependent initial phase \(\Bt_0(y)\) in the drive~\cite{Peng2018b}.

        Our starting point remains the coupled layer model. To give an example of a particular \(\vec{B}\) which gives \(C_2=1\) for the \(x=0\) edge state, we may take
        \begin{equation}
            B_0 = 3 - \sum_{j=1}^4 \cos \theta_j(t) , \quad B_{1 \leq j \leq 4} = \sin \theta_j(t)
        \end{equation}
        where the \(4\times 4\) \(\Gamma_j\) matrices may be expressed as tensor products of Pauli matrices \(\sigma^\alpha\) and \(\tau^\beta\):
        \begin{multline}
            \Gamma_0 = \tau^x,\;
            \Gamma_1 = \sigma^z\tau^z, \;
            \Gamma_2 = \sigma^x\tau^z, \\
            \Gamma_3 = \sigma^y\tau^z, \;\text{and}\;
            \Gamma_4 = \tau^y.
        \end{multline}
        The coupled layer model is then defined through the general construction above.

        To find a \((2+3)\)-dimensional model, we exchange one synthetic dimension for a spatial dimension~\cite{Long2021}. The mapping to accomplish this is provided by the frequency lattice description: Fourier modes of drive 1 are hopping terms in synthetic space, which, at a formal level, may be declared to be an actual spatial dimension. At the most direct level, the synthetic electric field \(\Omega_1 \hat{e}_1\) should be replaced by a linear potential in real space. Alternatively, a different form of spatial inhomogeneity may be introduced, provided that it causes localization and does not change the topological phase of the model.

        In more detail, the Hamiltonian for a \((1+D)\)-dimensional ALTP may be written as a sum of quadratic terms,
        \begin{equation}
            H(\Bt) = \sum_{x,x'} h^{\mu \nu}_{x x'}(\Bt) c^\dagger_{x \mu} c_{x' \nu}.
        \end{equation}
        The \(\theta_1\) Fourier components of \(h^{\mu \nu}_{x x'}(\Bt)\),
        \begin{equation}
            h^{\mu \nu}_{\vec{x} \vec{x}'}(\Bt_{\hat{1}}) = \frac{1}{2\pi}\int \d \theta_1\, h^{\mu \nu}_{x x'}(\Bt) e^{i(y-y')\theta_1},
        \end{equation}
        may be interpreted as hopping matrices. Here, \(\vec{x} = x\hat{x} + y\hat{y}\), \(y-y'\) indexes the Fourier component, and \(\Bt_{\hat{1}} = \sum_{j=2}^D \theta_j \hat{e}_j\) is \(\Bt\) with the \(\theta_1\) component removed. A Hamiltonian for a \((2+(D-1))\)-dimensional ALTP is then
        \begin{equation}
            H'(\Bt_{\hat{1}}) = H_{\mathrm{dis}}' + \sum_{\vec{x},\vec{x}'} h^{\mu \nu}_{\vec{x} \vec{x}'}(\Bt_{\hat{1}}) c^\dagger_{\vec{x} \mu} c_{\vec{x}' \nu},
        \end{equation}
        where \(H'_{\mathrm{dis}}\) includes a disorder potential in the \(y\) dimension (a linear potential, or otherwise). In this construction, the quasiperiodic hopping coefficients \(h^{\mu \nu}_{\vec{x} \vec{x}'}\) decay exponentially in \(|y-y'|\), but are only strictly local in the \(x\) dimension.

        In a strip geometry for \(H'\), with \(0 \leq 1-\epsilon \ll 1/2\), \(C_2\) edge states exist at the one-dimensional boundaries parallel to \(y\) (\autoref{fig:4DQHE}). In the frequency lattice, there is a linear potential (electric field) along \(\BO_{\hat{1}} = \sum_{j=2}^D \Omega_j \hat{e}_j\). To observe the quadratic response of a four-dimensional quantum Hall state, we must have a way of introducing a magnetic field through a plane which includes two of \((\hat{y}, \hat{e}_2,\hat{e}_3,\hat{e}_4)\).

        In fact, Ref.~\cite{Peng2018b} has already demonstrated how this may be done in a \((1+3)\)-dimensional wire model (where \(C_2 \neq 0\) requires adiabaticity or fine tuning). The initial phase \(\Bt_0\) appears in the frequency lattice as a vector potential. Including a spatially varying initial phase introduces a nonzero flux in short loops in the frequency lattice. Taking
        \begin{equation}
            \Bt_0 = B y \hat{e}_2
        \end{equation}
        introduces a flux \(B\) through each square plaquette in the \((\hat{y},\hat{e}_2)\)-plane. (Physical response only depends on \(B \mod 2\pi\).)

        By analogy to the (continuum) response of a four-dimensional Hall insulator, the average energy current between the \(\theta_3\) and \(\theta_4\) drives is found to be~\cite{Peng2018b}
        \begin{equation}
            \bar{P} = \frac{\Omega_3 \Omega_4}{(2\pi)^2} B L_y C_2 + O(B^2),
        \end{equation}
        where \(L_y\) is the length of the pumping boundary. This agrees with Eq.~\eqref{eqn:P4DQHE_intro}, as \(C_2 = W\). Observing pumping requires, as usual, filling a distance \(s \gg \zeta\) from the boundary with fermions~\footnote{There may be subextensive corrections from the corners of the sample, which can be canceled by driving the corners~\cite{Peng2018b}.}.

\section{Discussion}
    \label{sec:disc}

    Toy models capturing the physics of a system are essential in the study of complicated effects~\cite{Su1979,Haldane1988,Rudner2013}. The coupled layer construction for the QP pump allows for the straightforward identification of localized pumping modes at the edge, and an improved understanding of the topological-trivial transition. Similar models of ALTPs with more drives enable a comparably straightforward analysis of the edge modes, including the synthetic four-dimensional quantum Hall response of the \((2+3)\)-dimensional ALTP.

    With these models in hand, a systematic study of the observable responses of higher \(D\) ALTPs can be made. This is crucial when seeking technological applications of ALTPs, and a necessary ingredient for a more complete theoretical understanding of these phases. Some applications of ALTP phenomenology are known---energy pumping can be used to prepare highly excited nonclassical cavity states~\cite{Long2022b}---but finding ways to exploit other behaviors of ALTPs remains an interesting opening for future research.

    In a similar direction, finding experimentally feasible models for ALTPs would be very useful. The coupled layer models serve as a theoretical toy---they are not obviously suitable for realization in the laboratory. However, they could serve as a guide towards what features are necessary in an experimental Hamiltonian. The effect of dissipation and decoherence (as occurs in any experimental realization) on the energy pumping response should also be considered.

    Our analysis has been focused on edge modes, but ALTPs also possess a quantized bulk response~\cite{Long2021,Nathan2020b}. In the QP pump, the bulk energy circulation can be understood qualitatively through the coupled layer model (\autoref{fig:wires}). Extending this qualitative understanding to a quantitative one could provide access to bulk observables in higher \(D\) ALTPs, and potentially illuminate the nature of the bulk-edge correspondence in these phases.

    Our analysis of the QP pump transition was consistent with the two-dimensional quantum Hall transition universality class~\cite{Khmelnitskii1983,Levine1983,Laughlin1984,Chalker1988,Lee1992,Kivelson1992,Slevin2009,Puschmann2019}. However, our finite size scaling does not fix the critical exponents with high precision, nor is the theoretical connection concrete. A more extensive study of this transition may provide confirmation of this conjectured universality class. Due to the similarity of the coupled layer model to an SSH model in a rotating frame, it may be amenable to a real-space renormalization group analysis~\cite{Fisher1995,Fisher1999,Refael2013}. Alternatively, a network model for the transition, similar to the Chalker-Coddington model, would make the connection to quantum Hall systems transparent~\cite{Chalker1988}.

    The topological pumping response of the QP pump persists for an extremely long time when weak interactions are introduced. However, we have not observed a regime where the coupled layer model is asymptotically localized. Given that current numerical studies of MBL can no longer confirm such a phase in static systems~\cite{Abanin2021,Suntajs2020,Schulz2020,Taylor2021,Sels2021,Sels2021b,Morningstar2022,Sels2021a,Sierant2022}, it seems unlikely that a larger numerical study will be able to observe such a regime in quasiperiodically driven systems. Instead, it would be interesting to better understand the mechanism of thermalization in the accessible parameter regime. In static systems, this is likely due to many-body resonances between macroscopically distinct states~\cite{Gopalakrishnan2015,Khemani2017,Villalonga2020,Crowley2020b,Morningstar2022,Garratt2021,Garratt2022,Long2022c}. Extending the improved understanding of such resonances in static systems to the quasiperiodically driven setting would be interesting.

\section*{Acknowledgements}

    The authors thank O. Balabanov, C. Chamon, M. Hermanns, M. Kolodrubetz, C. Laumann, T. Neupert, C. Ortega-Taberner, and D. Vuina for helpful discussions. This work was supported by: NSF Grant No. DMR-1752759, and AFOSR Grant No. FA9550-20-1-0235 (D.L. and A.C.); and the NSF STC ``Center for Integrated Quantum Materials'' under Cooperative Agreement No. DMR-1231319 (P.C.). Numerical work was performed on the BU Shared Computing Cluster, using \textsc{quspin}~\cite{Weinberg2017,Weinberg2019}.

\appendix

\section{Frequency Lattice Localization}
    \label{app:frq_lat_loc}

    Anomalous localized topological phases (ALTPs) are localized in real space, but also in the synthetic dimensions. This is a crucial feature that prevents them from forming a featureless state at long times. The numerics reported in \autoref{sec:numerics} and \autoref{sec:interactions} use carefully chosen values of the model parameters for which states are well localized in both the spatial and synthetic dimensions.

    To numerically quantify the extent of eigenstates in both the spatial and synthetic dimensions we used an \emph{average spectral entropy}, \(H[S](t)\)~\cite{Crowley2019}. This does not require us to solve for the quasienergy states---the definition only depends on the values of correlation functions.

    \begin{figure}
        \centering
        \includegraphics[width=\linewidth]{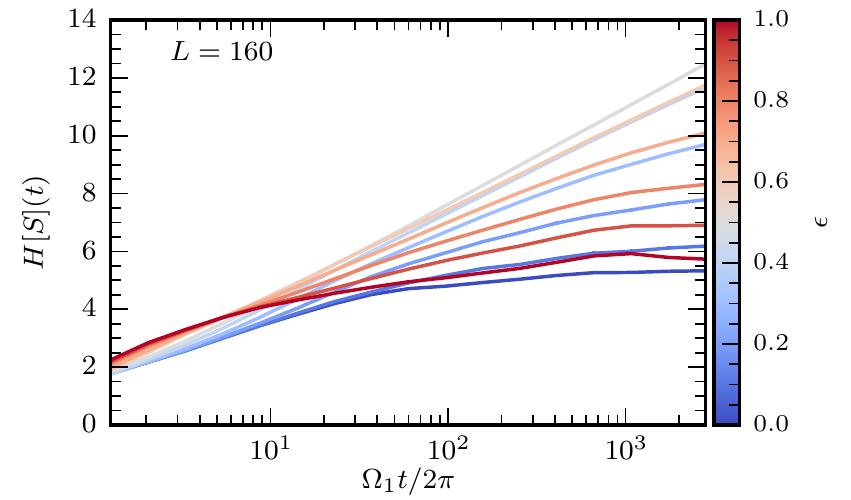}
        \caption{The average spectral entropy \(H[S](t)\) grows logarithmically with the observation time \(t\) in a delocalized phase. Deep in the localized phase, away from \(\epsilon=1/2\), \(H[S](t)\) saturates at a finite value. The finite size scaling analysis of \autoref{subsec:crit_point} suggests that all \(H[S](t)\) curves except that for \(\epsilon =1/2\) will saturate. \emph{Parameters:} As in \autoref{fig:corr_len}, except \(\delta/\Omega_1 = 0.15\). The spacing \(\d t\) between time points used in \(C_{ab}(x,t)\) when computing the Fourier transform is \(\Omega_1 \d t = 0.1\). Error bars are typically smaller than the line width.}
        \label{fig:HSt}
    \end{figure}

    \begin{figure}
        \centering
        \includegraphics[width=\linewidth]{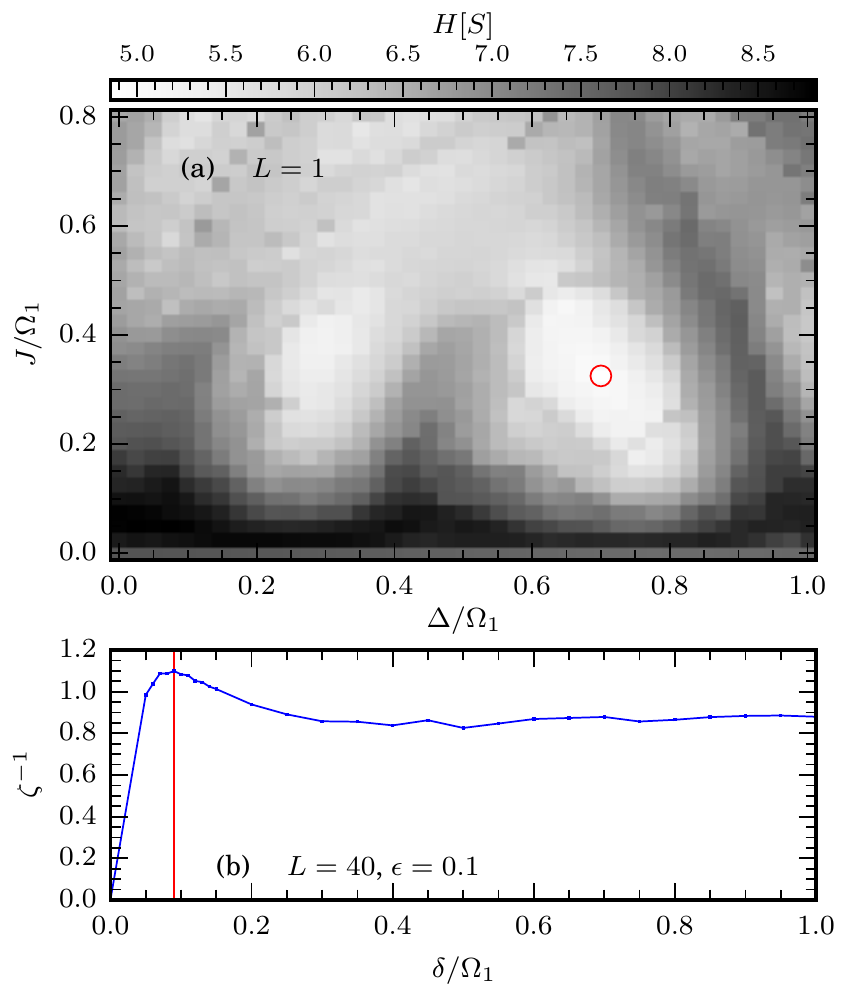}
        \caption{(\textbf{a}) Working parameters for \(J\) and \(\Delta\) are found by minimizing the average spectral entropy \(H[S]\) in the single-site problem. The optima are found to be \(J/\Omega_1 \approx 0.305\) and \(\Delta/\Omega_1 \approx 0.7\) (red circle). (\textbf{b}) The disorder strength \(\delta\) is subsequently optimized by minimizing the localization length (as measured in \autoref{fig:corr_len}) in an extended chain. The optimum is found to be \(\delta/\Omega_1 \approx 0.09\) (red line). Note that in no case is the localization behavior monotonic in these parameters. \emph{Parameters:} \(\epsilon=0.1\). (\textbf{a}) The maximum observation time for \(C_{\alpha\beta}\) is \(\Omega_1 t/2\pi \approx 2800\), \(L=1\), \(B_0 / \Omega_1 = 1\). (\textbf{b}) \(L = 40\), periodic boundary conditions.}
        \label{fig:params}
    \end{figure}

    First, define the single-particle states
    \begin{multline}
        \ket{z_0} = c^\dagger_{0 \uparrow} \ket{0}, \;
        \ket{x_0} = \tfrac{1}{\sqrt{2}}(c^\dagger_{0 \uparrow} + c^\dagger_{0 \downarrow}) \ket{0}, \\
        \text{and}\;
        \ket{y_0} = \tfrac{1}{\sqrt{2}}(c^\dagger_{0 \uparrow} + i c^\dagger_{0 \downarrow}) \ket{0},
    \end{multline}
    (with time evolved states \(\ket{\alpha_0(t)}\)) and the quadratic local observables
    \begin{equation}
        \Sigma^\alpha_x = \sigma^\alpha_{\mu\nu} c^\dagger_{x \mu} c_{x \nu}.
    \end{equation}
    Then the correlation functions
    \begin{equation}
        C_{\alpha \beta}(x,t) = \qexp{ \Sigma^\beta_x }{\alpha_0(t)}
    \end{equation}
    probe both the spatial extent of a particle initialized at position \(x = 0\) and its time dependent evolution within a single site. A large frequency lattice localization length is indicated in \(C_{\alpha \beta}(x,t)\) by quasiperiodic oscillations with significant weight in many harmonics.

    More precisely, the power spectrum
    \begin{equation}
        S_{\alpha \beta}(x,\omega) = |\mathcal{F}\{C_{\alpha \beta}\}(x, \omega)|^2
    \end{equation}
    will have support on many \(x\) and \(\omega\) if the frequency lattice extent is large (where \(\mathcal{F}\{\cdot\}\) is the Fourier transform with respect to time). On the other hand, if the quasienergy states are localized then \emph{all} power spectra \(S_{\alpha \beta}\) should only have significant weight on a few \(x\) and \(\omega\).

    To be sensitive to delocalization in any observable, we use the averaged power spectrum
    \begin{equation}
        S(x,\omega) = \frac{1}{9}\sum_{\alpha,\beta \in \{x,y,z\}} S_{\alpha \beta}(x,\omega).
    \end{equation}
    The extent of the support of \(S\) is quantified by its (Shannon) entropy,
    \begin{equation}
        H[S] = -\sum_{x, \omega} p(x,\omega) \ln p(x,\omega),
    \end{equation}
    where,
    \begin{equation}
        p(x, \omega) = S(x,\omega)/ \sum_{x',\omega'} S(x',\omega').
    \end{equation}
    
    Numerically, we can only compute \(S\) at finitely many points in \(x\) and \(\omega\). Delocalization is revealed by an unbounded growth of \(H[S]\) when the system size and integration time are increased. We will denote the average spectral entropy with an explicit time dependence, \(H[S](t)\), to emphasize this (\autoref{fig:HSt}). 
    
    Even a single site can have finite extent in the frequency lattice when driven quasiperiodically. To determine working values for the hopping strength \(J\)~\eqref{eqn:H_hop} (\(H_\mathrm{hop}\) has an on-site component which flips (\(\pm\)) pumping states to (\(\mp\)) states) and the on-site detuning \(\Delta\)~\eqref{eqn:H_dis}, we computed \(H[S](t)\) for a fine grid of values and a fixed large \(t\) (\autoref{fig:params}).

    Note that \(H[S]\) is not monotonic with either \(\Delta\) or \(J\). Indeed, taking \(J \to 0\) produces quasienergy states which have a Chern number, which must be delocalized. On the other hand, taking any energy scale much larger than \(|\BO|\) produces a model in the frequency lattice where the hopping terms are much larger than the inhomogeneity, which tends to delocalize. The optimal values of \(\Delta\) and \(J\) are both \(O(1)\):
    \begin{equation}
        J/\Omega_1 \approx 0.305, \quad \Delta/\Omega_1 \approx 0.7.
        \label{eqn:JDopt}
    \end{equation}

    Of course, the spatial disorder strength \(\delta\) (Eq.~\eqref{eqn:H_dis}) also controls the localization properties of the extended model with \(L>1\). With \(J\) and \(\Delta\) fixed as in Eq.~\eqref{eqn:JDopt}, we can subsequently find a working value for \(\delta\) by minimizing the localization length \(\zeta\), as computed in \autoref{fig:corr_len}. Again, the localization length is not a monotonic function of the disorder. The optimal value is found to be
    \begin{equation}
        \delta/\Omega_1 \approx 0.09.
    \end{equation}

    For these parameters, the spatial localization length is roughly one lattice site. This can be made smaller still by tuning \(\epsilon(1-\epsilon)\) closer to zero.

\bibliography{designer_ALTP}

\end{document}